\documentclass[10pt,twocolumn,letterpaper]{article}

\usepackage{cvpr}              

\usepackage[dvipsnames]{xcolor}

\definecolor{cvprblue}{rgb}{0.21,0.49,0.74}
\usepackage[pagebackref,breaklinks,colorlinks,citecolor=cvprblue]{hyperref}
\usepackage[hang,flushmargin]{footmisc}

\def\paperID{12249} 
\def\confName{CVPR}
\def\confYear{2024}

\title{Training Generative Image Super-Resolution Models by Wavelet-Domain Losses Enables Better Control of Artifacts\footnote{foot} \vspace{-10pt}}
\author{Cansu Korkmaz, A. Murat Tekalp, Zafer Dogan \\ 
College of Engineering and KUIS AI Center, Koc University\\
{\tt https://github.com/mandalinadagi/WGSR}
}

\begin{document}
\maketitle
\begin{abstract} \vspace{-10pt}
Super-resolution (SR) is an ill-posed inverse problem, where the size of the set of feasible solutions that are consistent with a given low-resolution image is very large. Many algorithms have been proposed to find a ``good" solution among the feasible solutions that strike a balance between fidelity and perceptual quality. Unfortunately, all known methods generate artifacts and hallucinations while trying to reconstruct high-frequency (HF) image details. A fundamental question is: Can a model learn to distinguish genuine image details from artifacts? Although some recent works focused on the differentiation of details and artifacts, this is a very challenging problem and a satisfactory solution is yet to be found. This paper shows that the characterization of genuine HF details versus artifacts can be better learned by training GAN-based SR models using wavelet-domain loss functions compared to RGB-domain or Fourier-space losses. Although wavelet-domain losses have been used in the literature before, they have not been used in the context of the SR task. More specifically, we train the discriminator only on the HF wavelet sub-bands instead of on RGB images and the generator is trained by a fidelity loss over wavelet subbands to make it sensitive to the scale and orientation of structures. Extensive experimental results demonstrate that our model achieves better perception-distortion trade-off according to multiple objective measures and visual evaluations.
\end{abstract}

\let\thefootnote\relax\footnotetext{* This work is supported by TUBITAK 2247-A~Award No. 120C156, TUBITAK 2232 Int. Fellowship for Outstanding Researchers Award No. 118C337,  KUIS AI Center, and Turkish Academy of Sciences~(TUBA).}    
\vspace{-10pt}
\section{Introduction}
\label{sec:intro}

\begin{figure}
\centering
\includegraphics[width=0.995\linewidth]{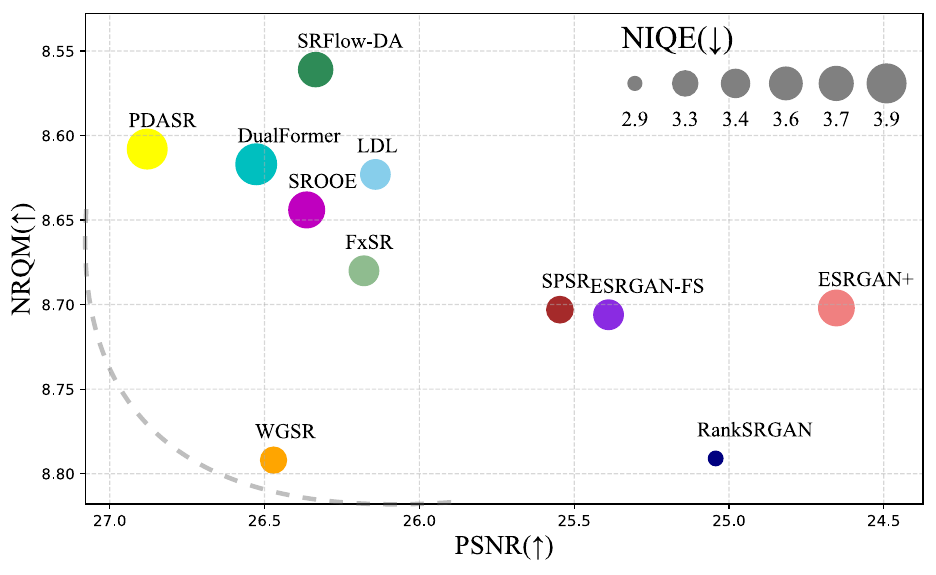} \vspace{-20pt}
\caption{Perception-distortion trade-off performance of our model WGSR vs. state-of-the-art methods on the PSNR-NRQM plane. Dashed curve shows the theoretical limit explained in \cite{Blau_2018}.}
\label{fig:pd_main_graph}
\vspace{11pt}
\end{figure}

\begin{figure}[!t]
\centering
\begin{subfigure}{0.155\textwidth}
\centering
\captionsetup{justification=centering}
    \begin{subfigure}{\textwidth}
        \includegraphics[width=\textwidth]{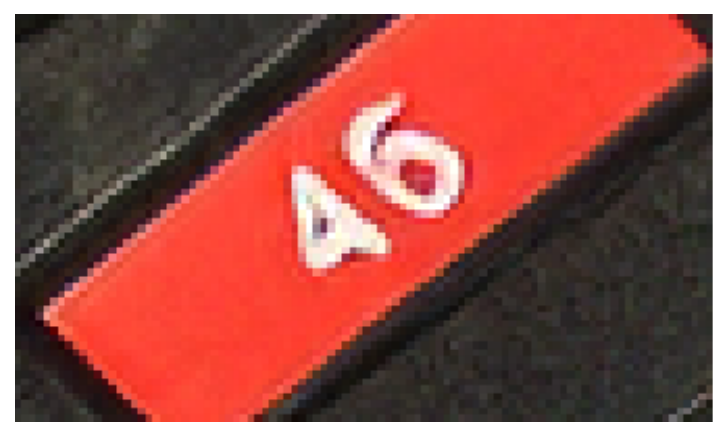}
    \end{subfigure} \vspace{-16pt}
    \caption*{ESRGAN-FS \cite{freq_sep} \\ (25.87 / 0.185)}
    \begin{subfigure}{\textwidth}
        \includegraphics[width=\textwidth]{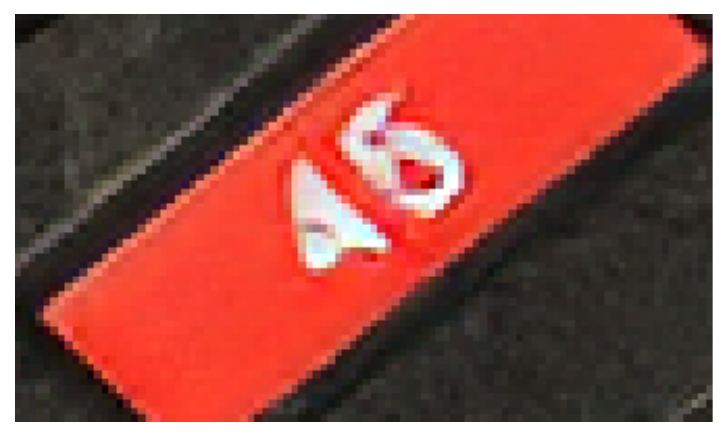}
    \end{subfigure} \vspace{-16pt}
    \caption*{RankSRGAN \cite{zhang2021ranksrgan} \\ (25.68 / 0.190)}
    \begin{subfigure}{\textwidth}
        \includegraphics[width=\textwidth]{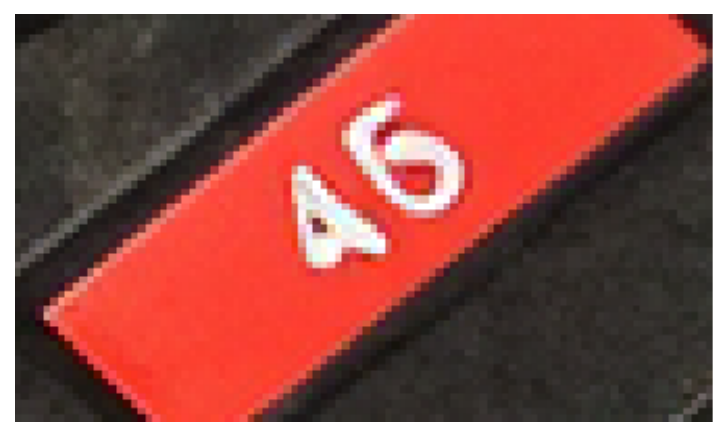}
    \end{subfigure} \vspace{-16pt}
    \caption*{LDL \cite{details_or_artifacts} \\ (26.67 / 0.194)}
\end{subfigure}
\begin{subfigure}{0.155\textwidth}
\centering
\captionsetup{justification=centering}
    \begin{subfigure}{\textwidth}
        \includegraphics[width=\textwidth]{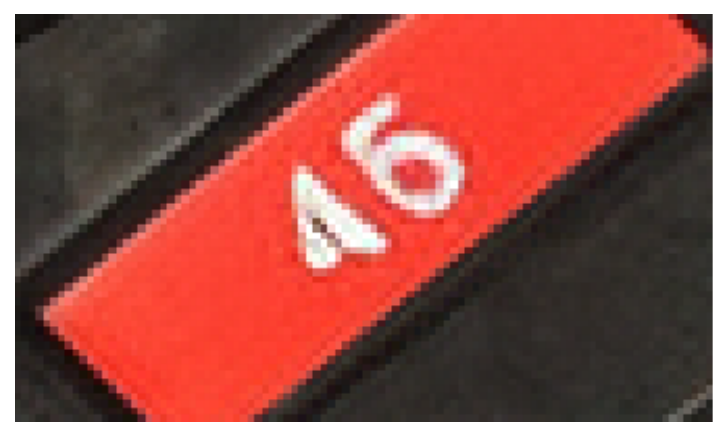}
    \end{subfigure} \vspace{-16pt}
    \caption*{PDASR \cite{PDASR} \\ (26.86 / 0.196)}
    \begin{subfigure}{\textwidth}
        \includegraphics[width=\textwidth]{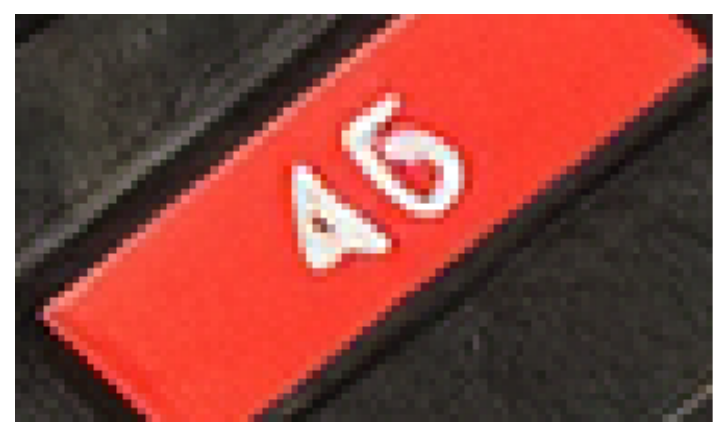}
    \end{subfigure} \vspace{-16pt}
    \caption*{FxSR \cite{fxsr} \\ (26.25 / 0.187)}
    \begin{subfigure}{\textwidth}
        \includegraphics[width=\textwidth]{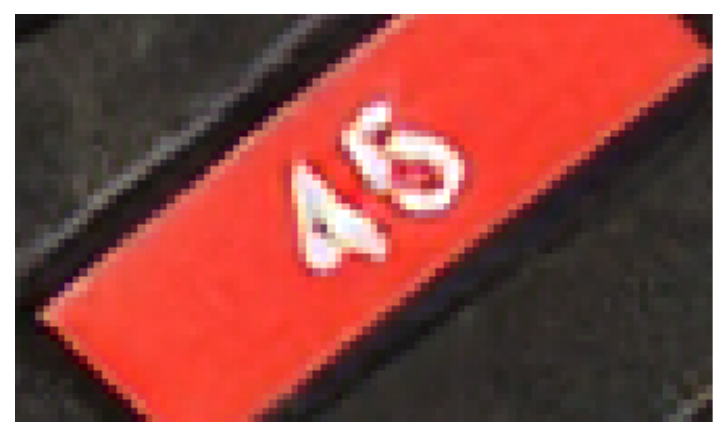}
    \end{subfigure} \vspace{-16pt}
    \caption*{WGSR (Ours) \\ (26.69 / 0.193)}
\end{subfigure}
\begin{subfigure}{0.155\textwidth}
\centering
\captionsetup{justification=centering}
    \begin{subfigure}{\textwidth}
        \includegraphics[width=\textwidth]{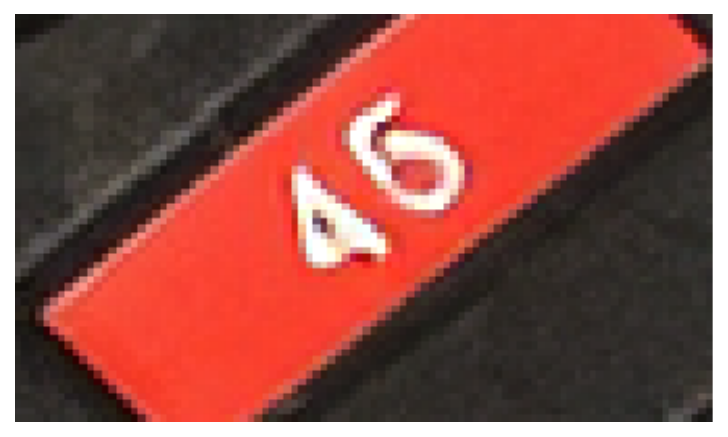}
    \end{subfigure} \vspace{-16pt}
    \caption*{DualFormer \cite{dualformer_luo2023effectiveness} \\ (26.98 / 0.188)}
    \begin{subfigure}{\textwidth}
        \includegraphics[width=\textwidth]{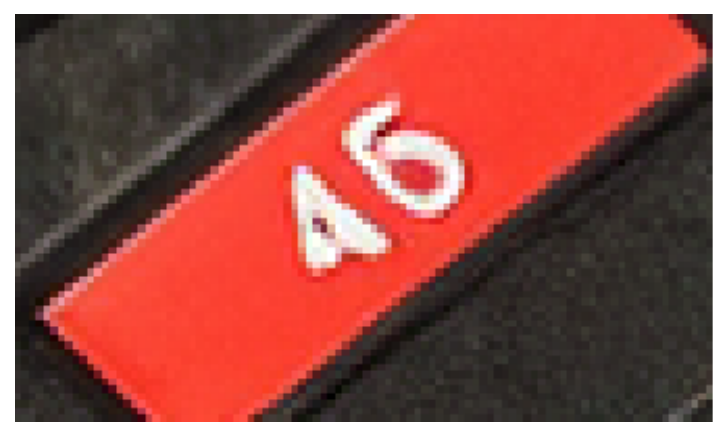}
    \end{subfigure} \vspace{-16pt}
    \caption*{SROOE \cite{srooe_Park_2023_CVPR} \\ (26.67 / 0.184)}
        \begin{subfigure}{\textwidth}
        \includegraphics[width=\textwidth]{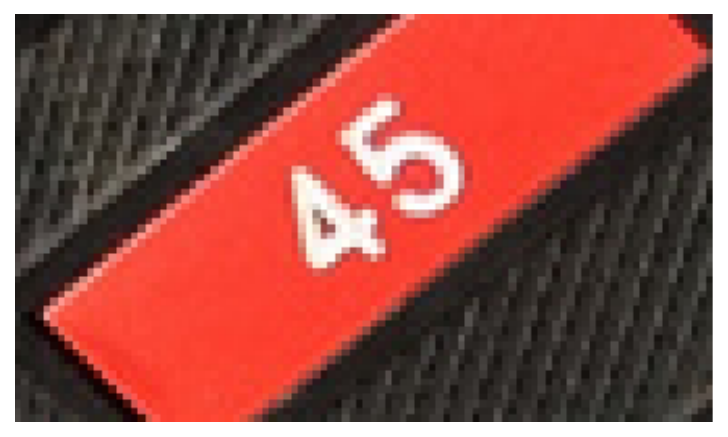}
    \end{subfigure} \vspace{-16pt}
    \caption*{HR \\ (PSNR$\uparrow$ / DISTS$\downarrow$)}
\end{subfigure} \vspace{-15pt}
\caption{Visual performance of recent $\times$4 SR methods on a~crop from Urban100 dataset (img-6) \cite{urban100_cite}. SOTA methods reconstruct ``5" as ``6", whereas the opening in the lower part of ``5" is visible in our result confirming that our model strikes a better balance between fidelity and visual quality. Note PSNR, DISTS and other quantitative scores are not good indicators of such artifacts.
}
\label{fig:first_img} 
\end{figure}

Single image super-resolution (SR) aims to reconstruct high-frequency (HF) details missing in low-resolution (LR) images. Early deep-learning based SR works employed simple convolutional neural networks (CNN), trained by pixel-wise $l_1$ and $l_2$ fidelity losses \cite{dong_srcnn, kim2016accurate}. They were followed by better models, which adopted residual \cite{ledig2017photorealistic, EDSR2017, RCAN2018} and dense connections~\cite{tong_densenet, zhang_res_dense}. Later, the spatial attention, channel attention~\cite{RCAN2018, dai_attention, niu_han, mei_nlsa, zhang2021tsan} and transformer networks \cite{Liang2021SwinIRIR, zhang_elan} have demonstrated impressive performance in terms of peak-signal-to-noise ratio (PSNR) and structural similarity measure (SSIM). However, minimization of mean-square error favors a probability-weighted average of all feasible SR outputs; hence, models that are optimized based only on fidelity losses produce overly smoothed images that lack HF details.

In order to generate visually more appealing results, generative SR models such as generative adversarial networks~(GANs) \cite{ledig2017photorealistic,wang2018esrgan, zhang2021ranksrgan, rottshaham2019singan, ma_SPSR}\cite{johnson2016perceptual, ledig2017photorealistic, lpips, Ma_NRQM, dists}, flow models \cite{srflow,ncsr}, and diffusion models \cite{compvis_rombach2022high, sr3_saharia2022image, gao2023implicit, wav_diff_Phung_2023_CVPR} have been proposed. Generative SR models aim to sample predicted SR images from a distribution that is similar to that of ground-truth (GT) images. However, they are known to hallucinate HF details and produce structural artifacts. Flow and diffusion models perform stochastic sampling in the sense that a single model can generate many samples. Hence, they allow less control per sample on learning details vs. artifacts. In this paper, we focus on conditional GAN-SR models, where a single trained model generates a single SR image sample. GAN models are trained by a weighted sum of pixel-wise fidelity and adversarial (discriminator) losses \cite{ian_gan}. Additional perceptual losses, such as the VGG loss \cite{ledig2017photorealistic}, the texture matching loss \cite{texture_mathing}, and the content loss \cite{Mechrez_contentloss} have been suggested to enforce feature-level similarity between SR and GT images to alleviate hallucinations and artifacts. However, perceptual losses are not sufficiently effective to control hallucinations and artifacts.

The perception-distortion (PD) trade-off hypothesis \cite{Blau_2018} states there is a bound beyond which any perceptual quality improvement (measured by a no-reference metric) comes at the expense of increased distortion (measured by a full-reference metric). Finding the best trade-off between fidelity and perceptual quality is not a well-defined optimization problem mainly because no quantitative perceptual image quality measure correlates well with human preferences. Recognizing this, recent SR challenges require consistency of SR reconstructions with the LR observations under the forward degradation model (also called feasible solutions) and conduct human evaluations for visual quality \cite{2021_ntire, Gu_2022_CVPR, 2022_ntire}. Yet, the size of the set of feasible solutions is very large, and determining which feasible solutions contain genuine image details and which contain artifacts or hallucinations is extremely challenging even for humans. 

In this paper, we propose a novel GAN-SR framework that uses wavelet-domain losses to suppress hallucinations and artifacts for a better PD trade-off. We define fidelity and adversarial losses over the subbands of the stationary wavelet transform~(SWT), where the scale and orientation of decomposed image features are well represented. Since the SWT decomposes an image without sub-sampling, it is able to provide the distinctive local features of low-frequency (LF) and HF subbands. Enforcing the reconstructed SR images to preserve local statistics within different subbands of HR images as an optimization goal enables the model to learn image details with different scales and orientations for a better PD trade-off.

Our wavelet-guided super-resolution (WGSR) model provides a better PD trade-off in the NRQM vs. PSNR plane compared to other state-of-the-art (SOTA) methods as shown in Fig.~\ref{fig:pd_main_graph}, where our NRQM score is the best among other methods with similar PSNR and our PSNR score is higher than RankSRGAN, which has similar NRQM score. Also, Fig. \ref{fig:first_img} demonstrates a visual comparison of our method and other SOTA methods. WGSR, shows remarkable performance by regulating easily visible artifacts, e.g., the opening in the lower part of ``5” is visible in our result, while other SOTA methods reconstruct ``5" as~``6." Note that quantitative scores, such as PSNR, DISTS and others, are not good indicators of such artifacts. To~the~best of our knowledge, our method is the first adversarial training scheme that employs wavelet guidance for artifact control, which can be applied to any GAN-SR model. To summarize, our primary contributions are: 
\begin{itemize}
    \item We propose a wavelet-domain fidelity loss (a weighted combination of $l_1$ losses on different wavelet sub-bands instead of the conventional RGB-domain $l_1$ loss), which is sensitive to the scale and orientation of local structures in images better observed in the SWT subbands.
    \item We propose utilizing an SWT-domain discriminator for adversarial training in order to control HF artifacts. We show that training the discriminator over HF wavelet subbands allows better control of the optimization landscape to segregate artifacts from genuine image details compared to the traditional RGB-domain discriminator. 
    \item We show that combining our proposed wavelet-guided training scheme with the RGB-domain DISTS perceptual loss (instead of the conventional VGG-based LPIPS loss) significantly improves fidelity (up to 0.5 dB in PSNR) with minimal (less than 1\%) loss in perceptual quality.
\end{itemize}

\section{Related Work}
\label{sec:formatting}

\begin{figure*}
\centering
\includegraphics[width=\linewidth]{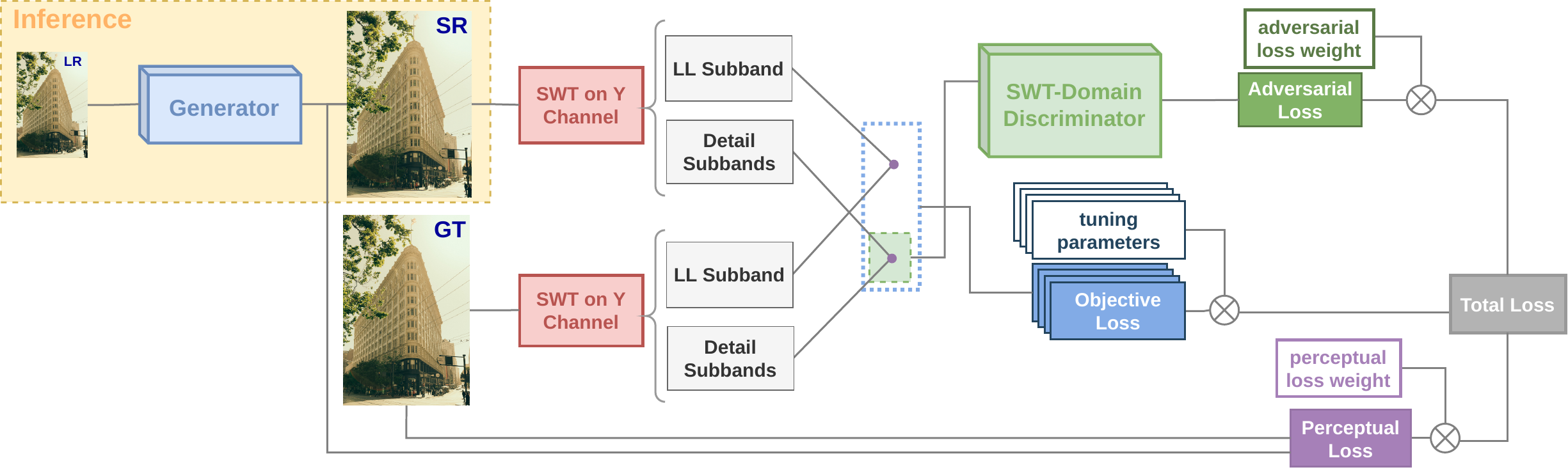} \vspace{-14pt}
\caption{Overview of the proposed GAN-SR framework guided by wavelet-domain losses, where the strength of the adversarial loss is tuned for each subband to control artifacts and the discriminator learns to decide whether the generated detail subbands are real or fake.}
\label{fig:arch}
\end{figure*}

\noindent
\textbf{GAN-based SR.}
\noindent
GANs \cite{ian_gan} offer a principled approach to achieve PD trade-off by controlling the weights of perceptual and fidelity losses in order to generate realistic images. To improve perceptual quality, Johnson \textit{et al.}~\cite{johnson2016perceptual} proposed a perceptual loss. Ledig \textit{et al.}~\cite{ledig2017photorealistic} proposed SRGAN with adversarial training along with the SRResNet generator. Wang \textit{et al.}~\cite{wang2018esrgan} proposed the ESRGAN with the Residual-in-Residual Dense Block (RRDB) architecture which has been employed as a standard backbone in many SOTA GAN-SR methods. Later, Rakotonirina \textit{et al.}~\cite{esrganplus} improved ESRGAN by additional noise injection and proposed ESRGAN+. Zhang \textit{et al.}~\cite{zhang2021ranksrgan} presented a Ranker that learns the behavior of perceptual metrics in RankSRGAN. Ma \textit{et al.}~\cite{ma_SPSR} proposed SPSR attenuate geometric distortions by preserving structure. Liang \textit{et al.}~\cite{details_or_artifacts} suggested a locally discriminative learning framework LDL by externally computing a probability map of each pixel being artifacts based on patch-level residual variances. Park \textit{et al.}~\cite{fxsr} introduced Flexible Style Image Super-Resolution (FxSR), which optimizes SR network with image-specific objectives without considering the regional characteristics. Later, in SROOE, Park \textit{et al.}~\cite{srooe_Park_2023_CVPR} proposed optimal objective estimation depending on perceptual and objective image maps. These methods \cite{details_or_artifacts, fxsr, srooe_Park_2023_CVPR} coexist with the computational burden of a large number of image maps. On the contrary, our wavelet-loss guided model does not require explicit calculation of an artifact map and inherently learns to suppress artifacts while retaining genuine details.  

\noindent \textbf{Training GANs by Frequency Domain Losses.}
Many studies have proposed frequency-related losses that better control the PD trade-off and ease the training of GANs~\cite{freq_sep, GuidedFreqSep, Fuoli2021FourierSL, gal2021swagan, schwarz2021frequency, ssd_gan_chen2021ssd}. Fritsche \textit{et al.}~\cite{freq_sep} proposed ESRGAN-FS, and the adversarial loss computed solely on the high-pass filtered images. Zhou \textit{et al.}~\cite{GuidedFreqSep} introduced CARB GAN-FS with two discriminator models that treat LF and HF components separately. In \cite{jiang2021focal}, Jiang \textit{et al.} proposed a focal frequency loss to alleviate the generation of frequency components that are hard to synthesize by means of a weighted Fourier space distance. Fuoli \textit{et al.}~\cite{Fuoli2021FourierSL} suggested Fourier domain discriminator to eliminate spectral discrepancies.  Recently, Luo \textit{et al.}~\cite{dualformer_luo2023effectiveness} introduced DualFormer, which utilizes spatial and spectral discriminators simultaneously. However, DFT domain (spectral) losses cannot localize HF image features according to scale and orientation, unlike wavelet decompositions, to characterize genuine details vs. artifacts.

\noindent
\textbf{Modeling SR in the Wavelet Domain.}
\noindent
Wavelet decomposition based approaches played crucial role in various computer vision tasks including GAN-inversion \cite{wagi_moon2023, ganinv_LIU2023286}, generative modeling \cite{gal2021swagan, wav_inpainting, wav_diff_Phung_2023_CVPR}, face-aging \cite{Liu2018AttributeAwareFA, awgan_2022}, video compression \cite{wang2020multi}, medical and thermal imaging \cite{wav_medical, wggan_ZHANG2021313}. 
Wavelet-domain learning methods have also been applied to SR tasks~\cite{DWSR_guo2017deep, wavelet_srnet_huang2017wavelet, WIDN_sahito2019wavelet, WRAN_xue2020wavelet, PDASR}; but existing methods directly predict wavelet coefficients of SR images. Specifically, Deng \textit{et al.} \cite{first_deng} proposed fusing images generated by objective and perceptual quality criteria via style transfer in the pixel domain. Later, Deng \textit{et al.}~\cite{Deng2019WaveletDS} employed Wavelet Domain Style Transfer (WDST), which performs style transfer on wavelet subbands. Zhang \textit{et al.}~\cite{PDASR} proposed PDASR to achieve PD trade-off by a two-stage SR framework that employs a low-frequency content constraint. PDASR reconstructs different frequency subbands independently, which causes inconsistency between subbands and results in unnatural artifacts. In contrast, we are the first to train pixel-domain ESRGAN \cite{wang2018esrgan} model using~weighted wavelet subband losses departing from~conventional RGB $\ell$1~loss. Our method, WGSR, is superior to others because predicting RGB pixels is easier than predicting sparse~wavelet coefficients of detail bands, while unequal weighting of losses in different wavelet subbands enables learning structures with different scales and orientations. To the best of our knowledge, WGSR is the first GAN-based RGB-domain SR model guided by wavelet-domain losses. 


\section{WGSR: Wavelet-Guided SR Framework}
\label{method}
We propose a novel adversarial training framework presented in Fig. \ref{fig:arch} for GAN-SR models that suppresses HF hallucinations and artifacts to achieve better PD trade-off by (i) training the discriminator only on the HF subbands, (ii) introducing a wavelet domain distortion loss to guide the generator, and (iii) selecting more suitable perceptual loss that couples better with our optimization objective. 

\subsection{Rationale for using Wavelet-Domain Losses}

The Stationary Wavelet Transform (SWT) allows multi-scale decomposition of images \cite{wavelet_doc} into one LF subband referred as LL and several HF (e.g., LH, HL, HH) subbands. The decomposition level of LL subband determines the number of HF subbands that convey the detailed information in horizontal, vertical, and diagonal directions, respectively. It is important to note that since the resolution is highly critical in SR tasks, we utilize the SWT rather than classical Discrete Wavelet Transform (DWT). The main difference of SWT is the removal of the decimation part in the DWT, hence, the SWT method inherently couples the scale/frequency information with spatial location. 

\begin{figure}[!t]
\centering
\begin{subfigure}{0.109\textwidth} 
    \begin{subfigure}{\textwidth}
        \includegraphics[width=\textwidth]{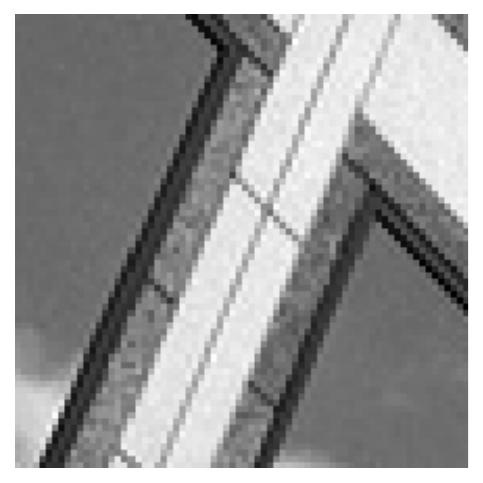}
        \scriptsize HR - PSNR
    \end{subfigure} 
    \begin{subfigure}{\textwidth}
        \includegraphics[width=\textwidth]{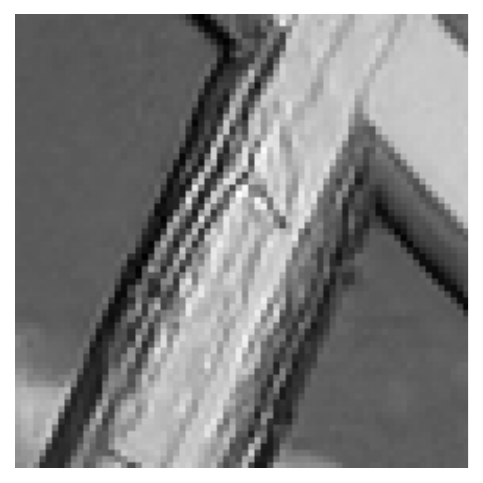}
        \scriptsize \cite{esrganplus} - 26.727 dB
    \end{subfigure} 
    \begin{subfigure}{\textwidth}
        \includegraphics[width=\textwidth]{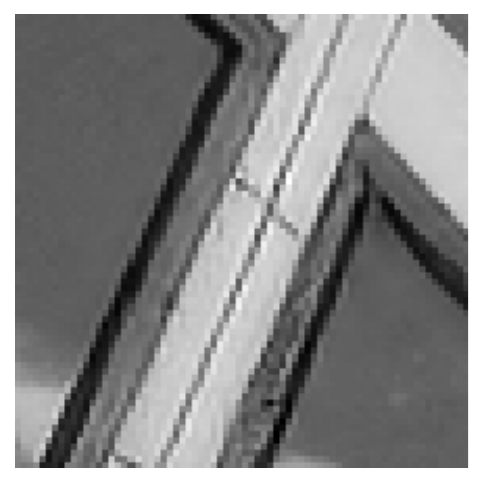}
        \scriptsize Ours - 31.239 dB
    \end{subfigure} 
\end{subfigure}
\begin{subfigure}{0.11\textwidth} 
    \begin{subfigure}{\textwidth}
        \includegraphics[width=\textwidth]{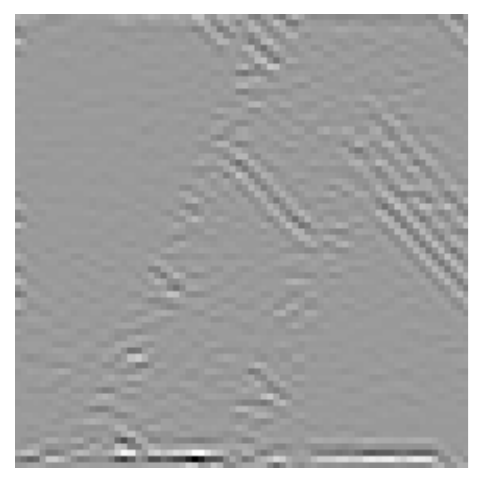}
        \scriptsize LH subband
    \end{subfigure} 
    \begin{subfigure}{\textwidth}
        \includegraphics[width=\textwidth]{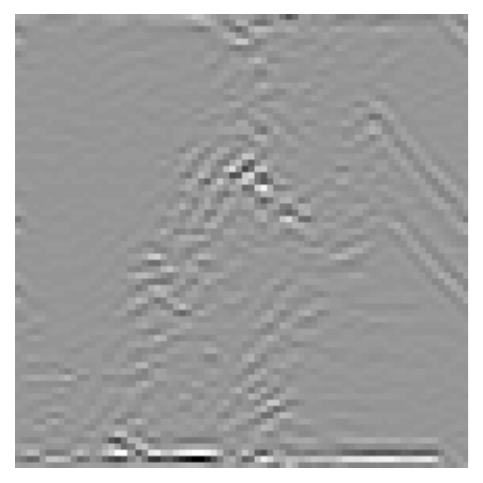}
        \scriptsize  33.033 dB
    \end{subfigure} 
    \begin{subfigure}{\textwidth}
        \includegraphics[width=\textwidth]{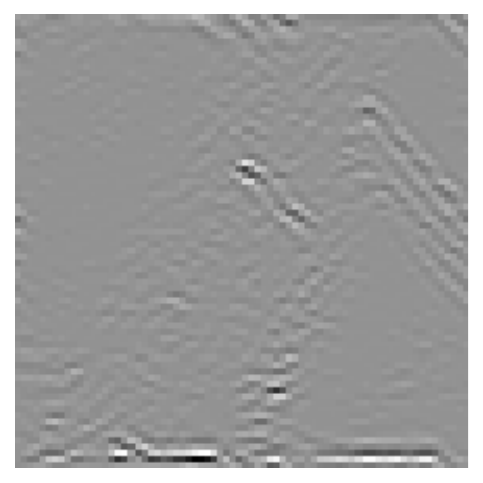}
        \scriptsize 34.491 dB
    \end{subfigure} 
\end{subfigure}
\begin{subfigure}{0.11\textwidth} 
    \begin{subfigure}{\textwidth}
        \includegraphics[width=\textwidth]{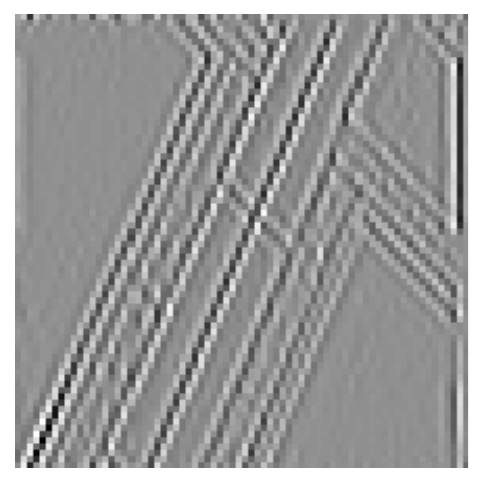}
        \scriptsize  HL subband
    \end{subfigure} 
    \begin{subfigure}{\textwidth}
        \includegraphics[width=\textwidth]{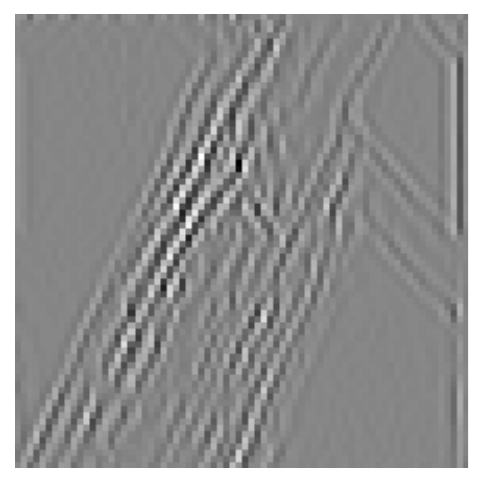}
        \scriptsize  22.983 dB
    \end{subfigure} 
    \begin{subfigure}{\textwidth}
        \includegraphics[width=\textwidth]{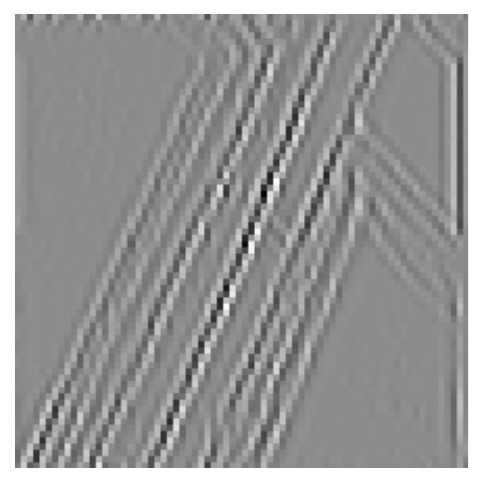}
        \scriptsize 26.367 dB
    \end{subfigure} 
\end{subfigure}
\begin{subfigure}{0.11\textwidth} 
    \begin{subfigure}{\textwidth}
        \includegraphics[width=\textwidth]{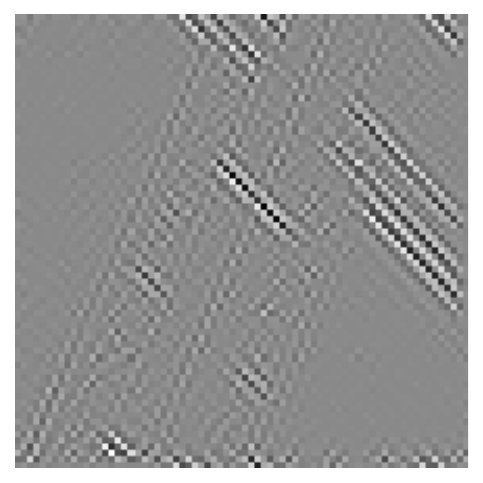}
        \scriptsize  HH subband
    \end{subfigure} 
    \begin{subfigure}{\textwidth}
        \includegraphics[width=\textwidth]{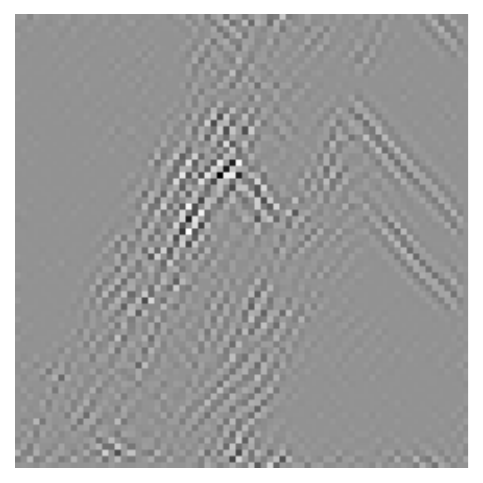}
        \scriptsize 28.098 dB
    \end{subfigure} 
    \begin{subfigure}{\textwidth}
        \includegraphics[width=\textwidth]{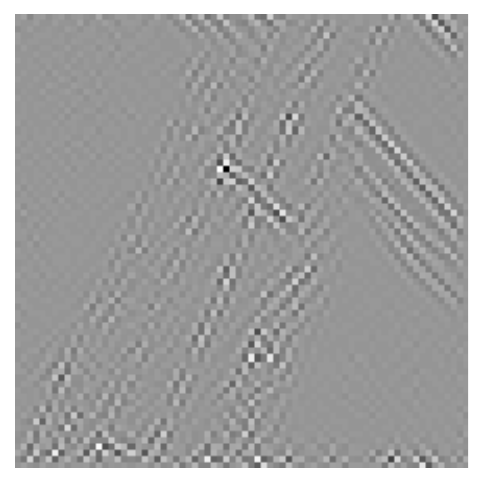}
        \scriptsize 29.971 dB
    \end{subfigure} 
\end{subfigure} \vspace{-4pt}
\caption{Illustration of our main premise that imposing different losses to different SWT subbands results in remarkable quantitative and qualitative performance improvements in GAN-based SR models. Specifically, enforcing fidelity loss on wavelet sub-bands instead of on RGB channels and running the discriminator only on detail (LH, HL, and HH) subbands helps eliminate visible artifacts caused by ESRGAN+ \cite{esrganplus} and leads to better preservation of details. Overall scores of our method WGSR (PSNR:~26.33/DISTS: 0.115) outperform ESRGAN+ \cite{esrganplus} (PSNR:~22.78/DISTS: 0.225).}
\label{fig:wavelet} 
\end{figure}

The LL subband of the SWT decomposition has a significant effect on the fidelity of the reconstructed images~\cite{Deng2019WaveletDS}. Hence, it is crucial not to alter the existing frequencies or introduce new ones into the LL subband to attain low distortion. At the same time, the HF contents of an image that are aligned with the LL spatial contents need to be reconstructed to achieve photo-realistic images. To better demonstrate the key advantages of SWT-guided adversarial training, we apply 1-level SWT decomposition to HR image, to the result of ESRGAN+ \cite{esrganplus} and to the result of our WGSR method and present these decompositions in Fig. \ref{fig:wavelet}. While training ESRGAN+ \cite{esrganplus}, there is no guidance provided by wavelet domain losses, hence it represents classical adversarial training approach, and the image generated by ESRGAN+ \cite{esrganplus} contains exaggerated artifacts. When we closely examine the HF subbands, due to the orientation of structures in the image, the HL subband contains more hallucination with higher distortion, resulting to have the lowest PSNR score among other subbands. So, this specific patch of the ESRGAN+ \cite{esrganplus} actually requires enhancement in the HL subband. However, extracting this information from the RGB image itself is much harder than doing it on the HL subband for the discriminator network and it fails to recognize this unnatural artifact in the vicinity of the window in the image. On the other hand, we optimize our discriminator network to separate the details from artifacts by only feeding the HF subbands as opposed to RGB images. As a result, our wavelet-guided optimization result shows significant improvement in all subbands, as well as in the final SR image; hence, it obtains remarkable photo-realistic result that contains genuine image details rather than hallucinated artifacts. 

\subsection{Architecture}
The proposed framework shown in Figure \ref{fig:arch} consists of an RGB-domain generator and a SWT-domain discriminator, which are jointly trained using SWT-guided fidelity and adversarial, and pixel-domain perceptual losses. The framework is generic in the sense that any generator and any discriminator model can be easily plugged into this framework.

\noindent
\textbf{SWT-domain Discriminator}
We employ a discriminator network that is tasked to learn how ``real" are the generated HF details (LH, HL, and HH subbands) compared to the ones that appear in the SWT decomposition of HR images. Our discriminator only evaluates the horizontal, vertical and diagonal details as opposed to evaluating RGB images, since they are crucial to control details vs. hallucinated artifacts. As shown in the last 3 columns of Fig. \ref{fig:wavelet}, LH, HL, and HH subbands convey sparse information, which simplifies the task of the discriminator and enables stable training. The training pipeline of the discriminator starts with YCbCr conversion of the generated image. The SWT decomposition is applied on the Y channel (Cb and Cr are discarded) to obtain LL, LH, HL and HH subbands. Only the details (LH,HL, HH) subbands are used to train the discriminator. The architecture of the discriminator consists of 9 convolution layers, whose kernel size alternates between 3x3 and 4x4, followed by 2D batch norms, and ReLU activation applied in between as in \cite{jolicoeur2018relativistic}. The number of output features of each convolutional layer increases from 64 to 512 and at the end, there are 2 linear layers with LeakyReLU activation which returns a 2D array to determine whether the HF subbands of the generated image resemble the ones of the GT image. Since this approach allows the discriminator to focus more on the relevant HF details of the generated images, which is where the artifacts are clearly separated from genuine details, it prevents hallucinations and eliminates distortions.

\noindent
\textbf{RGB-domain Generator}
The RRDB \cite{wang2018esrgan} architecture is selected as a backbone generator network, which consists of 23~residual-in-residual dense blocks without batch norm. Except for the output layer, all convolutional layers use 3x3 kernels with 64 features, and Leaky ReLU is selected as the activation function. Since the generator network takes randomly cropped RGB patches during training we refer to it as RGB-domain generator. It is worth mentioning that our proposed training scheme with wavelet domain losses and SWT-domain discriminator can be coupled with any generator network architecture. 

\subsection{Training by SWT-Domain Losses}
\label{sec:optim}
Instead of using the regular RGB-domain fidelity loss as in conventional GAN-SR methods, we define the SWT-domain fidelity loss, $L_{SWT}$, with corresponding tuning parameters. The flexibility of weighting the contribution of each subband individually enables adjusting the balance of fidelity and perceptual quality of the generated SR image. We sum the $l_1$ fidelity loss between SWT subbands of generated images $x$ and the GT image $y$ and average over a minibatch size denoted as $\mathbb{E}[.]$, given by 
\small
\begin{multline}  \label{eq:swt_fidelity}
 L_{SWT} = \mathbb{E} \bigl[ \sum_{j} \lambda_{j} \big\| {SWT(G(x))}_{j} - {SWT(y)}_{j} \big\| _1  \bigr]   
\end{multline} 
\normalsize 
where $G$ denotes the generator model and $\lambda_j$ are appropriate scaling factors to control the~generated HF details to avoid hallucinations and disturbing visual artifacts appearing around fine-scale regular structures such as sharp lines/edges on windows, buildings, letters, or tree branches.

\noindent
When the lowest frequency (LL) subband contains flat regions or large-scale structures, it is important to preserve the shapes of objects to maintain the objective quality. So, we~compute the adversarial loss term, given by equation \ref{eq:loss_adv_gan}, over the detail subbands (LH, HL, and HH) in order not to alter the existing frequencies or introduce new ones. 
\small
\begin{multline}\label{eq:loss_adv_gan}
 L_{adv,G} = - \mathbb{E} \bigl[ log (1 -D( {SWT(y)}_{*})) \bigr] \\ - \mathbb{E} \bigl[ logD ( {SWT(G(x))}_{*}) \bigr]  
\end{multline} 
\normalsize
where $D$ is the discriminator model, and $*$ indicates concatenation of details subbands.

\noindent
Then, the overall loss for the generator is given by 
\begin{multline} \label{eq:loss_generator}
 L_{G} =  L_{SWT} + \lambda_{adv} \cdot L_{adv,G} + \lambda_{perc} \cdot L_{perc}
\end{multline}
where $L_{perc}$ denotes the perceptual loss, measuring errors in the feature space provided by image quality assessment DISTS \cite{dists}.

\noindent
The loss term for the discriminator, which only takes the HF details subbands as input, is given by
\small
\begin{multline} \label{eq:loss_disc}
 L_{D} = - \mathbb{E} \bigl[ logD( {SWT(y)}_{*}) \bigr]\\
  -  \mathbb{E} \bigl[ log ( 1 - D( {SWT(G(x))}_{*})) \bigr]
\end{multline}
\normalsize  

\noindent
Since determination of the optimal values $\lambda$ for each subband is not straightforward, we find the best PD trade-off point by searching empirically to be at $\lambda_{LL}$ = 0.1, $\lambda_{LH}$ = $\lambda_{HL}$ = 0.01, $\lambda_{HH}$ = 0.05, $\lambda_{adv}$ = 0.005 and $\lambda_{perc}$ = 1.

\begin{table*}
 \caption{Quantitative comparison of the proposed wavelet decomposition-based optimization objective vs. other state-of-the-art methods for $\times$4 SR task. The best and the second-best are marked in \textbf{bold} and \underline{underlined}, respectively.} \vspace{-3pt}
    \centering
    \scalebox{0.67}{
    \begin{tabular}{cccccccccccccc}
    \specialrule{.1em}{.05em}{.05em} 
    Benchmark & Metric & ESRGAN-FS & ESRGAN+ & SPSR & RankSRGAN & SRFlow-DA & LDL & FxSR & PDASR & DualFormer & SROOE & WGSR  & WGSR \\  
    & & \cite{freq_sep} & \cite{esrganplus} & \cite{ma_SPSR} & \cite{zhang2021ranksrgan} & \cite{jo2021srflowda} & \cite{details_or_artifacts} & \cite{fxsr} & \cite{PDASR} & \cite{dualformer_luo2023effectiveness} & \cite{srooe_Park_2023_CVPR} & (1-lvl) & (2-lvl)\\ 
    \hline
    Dataset &  & DF2K & DIV2K & DIV2K & DIV2K & DF2K & DIV2K & DIV2K & DIV2K & DIV2K & DF2K & DIV2K & DIV2K \\ \hline\hline
& PSNR $\uparrow$ & 30.329 & 29.002 & 30.357 & 28.859 & 30.764 & 30.964 & 30.858 & \textbf{31.728} & 31.299 & 31.285 & 31.334 & \underline{31.508}\\
& SSIM $\uparrow$ & 0.844 & 0.801 & 0.843 & 0.823 & 0.855 & 0.860 & 0.855 & \textbf{0.875} & \underline{0.869} & 0.867 & 0.866 & \underline{0.869}\\
& LPIPS $\downarrow$& 0.065 & 0.100 & 0.065 & 0.077 & 0.084 & 0.066 & \textbf{0.060} & 0.078 & 0.066 & \underline{0.064} & 0.065 & 0.068\\
& LPIPS-VGG $\downarrow$& 0.172 & 0.224 & 0.167 & 0.194 & 0.189 & 0.151 & 0.163 & 0.178& \underline{0.149} & \textbf{0.148} & 0.168 & 0.162\\
Set5 & DISTS $\downarrow$& 0.096 & 0.126 & \textbf{0.092} & 0.109 & 0.110 & \textbf{0.092} & 0.096 & 0.110 & \underline{0.093} & 0.094 & 0.110 & 0.107\\
& NIQE $\downarrow$& 4.320 & 4.710 & 4.215 & \textbf{3.589} & 5.600 & 4.602 & 4.642 & 5.520 & 5.225 &5.067 & \underline{4.175} & 4.270\\
& NRQM $\uparrow$& 8.015 & 8.250 & 8.079 & \textbf{8.613} & 6.886 & 7.588 & 7.973 & 7.416 & 7.156 & 7.428 & \underline{8.252} & 7.927\\
& PI $\downarrow$& 3.385 & 3.380 & 3.337 & \textbf{2.663} & 4.439 & 3.681 & 3.565 & 4.286& 4.210 & 3.884 & \underline{3.160} & 3.239\\ 
& LR-PSNR $\uparrow$& 42.950& 42.860& 43.630& 38.210& 49.940& 46.590& 50.210& \textbf{53.280}& 42.929 & \underline{51.020} & 49.911 & 50.868 \\ \hline
& PSNR $\uparrow$& 26.415 & 25.923 & 26.564 & 25.797 & 27.123 & 27.096 & 27.115 & \textbf{27.869}& 27.394 & 27.278 & \underline{27.395} & 26.689\\ 
 & SSIM $\uparrow$& 0.711 & 0.685 & 0.714 & 0.686 & 0.728 & 0.735 & 0.733 & \textbf{0.751} & \underline{0.739} & \underline{0.739} & \underline{0.739} & 0.716\\
 & LPIPS $\downarrow$& 0.140 & 0.159 & 0.132 & 0.145 & 0.133 & 0.131 & 0.123 & 0.142 & \underline{0.120} &\textbf{0.116} & 0.138 & 0.140 \\
 & LPIPS-VGG $\downarrow$& 0.241 & 0.274 & 0.237 & 0.261 & 0.254 & 0.225 & 0.227 & 0.247 & \underline{0.216} & \textbf{0.215} & 0.252 & 0.257\\
 Set14 & DISTS $\downarrow$ & 0.102 & 0.126 & 0.098 & 0.112 & 0.113 & 0.098 & 0.097 & 0.112& \underline{0.092} & \textbf{0.090} & 0.112 & 0.110\\
 & NIQE $\downarrow$& 3.586 & 3.495 & 3.657 & \textbf{3.220} & 4.238 & 3.635 & 3.578 & 4.109& 4.167 & 3.984 & 3.594 & \underline{3.309}\\
 & NRQM $\uparrow$& 8.029 & 8.046 & 8.056 & \textbf{8.227} & 7.843 & 7.907 & 7.992 & 7.818& 7.821 & 7.928 & 7.930 & \underline{8.196}\\
 & PI $\downarrow$& 2.838 & 2.768 & 2.908 & \textbf{2.519} & 3.196 & 2.959 & 2.880 & 3.251& 3.272 & 3.093 & 2.914 & \underline{2.577} \\ 
 & LR-PSNR $\uparrow$& 40.930 & 41.270& 41.390& 37.180& 49.570& 44.500& 49.000& \textbf{50.510}& 41.678 & \underline{49.150} & 49.023 & 48.937\\ \hline
 & PSNR $\uparrow$& 25.389 & 24.653 & 25.546 & 25.043 & 26.335 & 26.142 & 26.179 & \textbf{26.879} & \underline{26.527} & 26.364 & 26.471 & 26.372\\
 & SSIM $\uparrow$& 0.658 & 0.614 & 0.659 & 0.639 & 0.684 & 0.682 & 0.685 & \textbf{0.703} & 0.691 & 0.693 & \underline{0.696} & 0.684\\
 & LPIPS $\downarrow$& 0.166 & 0.211 & 0.161 & 0.183 & 0.191 & 0.163 & \underline{0.157} & 0.187 & 0.158 & \textbf{0.153} & 0.187 & 0.174\\
 & LPIPS-VGG $\downarrow$& 0.269 & 0.313 & 0.263 & 0.285 & 0.286 & 0.244 & 0.253 & 0.272 & \underline{0.242} & \textbf{0.241} & 0.283 & 0.282\\
 BSD100 & DISTS $\downarrow$& 0.119 & 0.151 & \underline{0.118} & 0.129 & 0.145 & \underline{0.118} & \underline{0.118} & 0.136 & 0.119 & \textbf{0.116} & 0.137 & 0.132\\
 & NIQE $\downarrow$& 3.386 & 3.675 & 3.261 & \textbf{2.903} & 3.603 & 3.383 & 3.386 & 3.902 & 3.957 & 3.684 & 3.428 & \underline{3.243}\\
 & NRQM $\uparrow$& 8.706 & 8.702 & 8.703 & 8.791 & 8.561 & 8.623 & 8.680 & 8.608 & 8.617 & 8.644 & \underline{8.792} & \textbf{8.793}\\
 & PI $\downarrow$& 2.402 & 2.531 & 2.335 & 2.086 & 2.631 & 2.473 & 2.422 & 2.779 & 2.796 & 2.576 &\textbf{2.053} & \underline{2.065}\\ 
 & LR-PSNR $\uparrow$& 39.910& 41.530& 40.990& 37.510& \textbf{49.920}& 43.690& 49.260& \underline{49.830}& 42.306 &49.610 & 49.046& 48.915\\ \hline
 & PSNR $\uparrow$& 24.556 & 23.235 & 24.795 & 24.121 & 25.632 & 25.491 & 25.668 & \textbf{26.279}& 25.686  & \underline{25.939} & 25.779 & 25.606 \\
 & SSIM $\uparrow$& 0.743 & 0.707 & 0.747 & 0.719 & 0.763 & 0.767 & 0.772 & \textbf{0.785} & 0.773 &  0.779 & \underline{0.781} & 0.777\\
 & LPIPS $\downarrow$& 0.124 & 0.143 & 0.119 & 0.143 & 0.129 & \underline{0.110} & \textbf{0.109} & 0.123 & 0.115 & 0.108 & 0.135 & 0.135\\
 & LPIPS-VGG $\downarrow$& 0.222 & 0.248 & 0.216 & 0.249 & 0.241 & \textbf{0.197} & 0.204 & 0.223 & 0.200 & \underline{0.199} & 0.243 & 0.243\\
 Urban100 & DISTS $\downarrow$& 0.090 & 0.104 & \underline{0.085} & 0.106 & 0.115 & \textbf{0.082} & 0.087 & 0.102 & \underline{0.085} & \underline{0.085} & 0.108 & 0.101\\
 & NIQE $\downarrow$ & 3.803 & 3.639 & 3.686 & 3.712 & 4.361 & 3.777 & 3.801 & 4.012 & 4.148 & 3.906 & \underline{3.526} & \textbf{3.326}\\
 & NRQM $\uparrow$ & 6.652 & 6.571 & 6.631 & 6.756 & 6.479 & 6.582 & 6.608 & 6.540 & 6.518 & 6.552 & \underline{6.827} & \textbf{7.406}\\
 & PI $\downarrow$& 3.590 & 3.562 & 3.549 & 3.278 & 3.918 & 3.617 & 3.603 & 3.750 & 3.831 & 3.695 & \underline{3.266} & \textbf{3.112}\\ 
 & LR-PSNR $\uparrow$&40.170& 39.200& 40.420& 36.390& \underline{49.790} & 44.570& 48.310&\textbf{50.900}& 41.367 & 48.570& 48.250& 48.125\\ \hline
 & PSNR $\uparrow$ & 28.073 & 26.770 & 28.190 & 27.196 & 28.954 & 28.959 & 29.022 & \underline{29.707} & 29.250 & 29.312 & 29.188 & \textbf{29.857} \\
 & SSIM $\uparrow$ & 0.770 & 0.743 & 0.772 & 0.740 & 0.789 & 0.795 & 0.798 & \underline{0.810} & 0.802 & 0.803 & 0.804 &\textbf{0.820} \\
 & LPIPS $\downarrow$ & 0.116 & 0.133 & 0.110 & 0.145 & 0.123 & \underline{0.101} & 0.103 & 0.123 & \textbf{0.097} & 0.103 & 0.104 & 0.111 \\
 & LPIPS-VGG $\downarrow$ & 0.226 & 0.242 & 0.218 & 0.250 & 0.253 & \textbf{0.199} & 0.212 & 0.237 & 0.202 & \underline{0.200} & 0.210 & 0.213 \\
 DIV2K & DISTS $\downarrow$ & 0.058 & 0.067 & 0.055 & 0.067 & 0.075 & \textbf{0.053} & 0.057 & 0.076 & 0.056 & \underline{0.054} & \underline{0.054} & 0.056 \\
 & NIQE $\downarrow$ & 2.953 & 2.911 & 2.952 & \textbf{2.576} & 3.828 & 2.966 & 3.064 & 3.439 & 3.237 & 3.464 & \underline{2.888} & 2.943 \\
 & NRQM $\uparrow$ & 6.724 & 6.721 & 6.694 & \underline{6.828} & 6.519 & 6.610 & 6.671 & 6.560 & 6.611 & 6.543 & \textbf{6.870} & 6.452 \\
 & PI $\downarrow$ & 3.137 & 3.126 & 3.158 & \textbf{2.891} & 3.650 & 3.213 & 3.231 & 3.462 & 3.340 & 3.516 & \underline{3.107} & 3.602 \\
 & LR-PSNR $\uparrow$ & 42.915 & 38.407 & 42.565 & 37.758 & 50.151 & 45.900 & 50.514 & \textbf{51.690} & \underline{51.088} & 42.950 & 49.057 & 49.076 \\
\specialrule{.1em}{.05em}{.05em} 
    \end{tabular} 
}
\label{table:quantitative_results}
\end{table*}

\section{Experiments}
\subsection{Experimental Setup}
\textbf{Training Details.}
As a training set, we used 800 LR images from DIV2K \cite{Agustsson_2017_CVPR_Workshops} that are generated using the MATLAB bicubic downsampling kernel with a scaling factor of 4$\times$. Randomly cropped 32$\times$32 pixels of RGB LR patches on a minibatch of 16 are given to the generator. Then the loss terms are calculated after applying SWT to the Y-channel of generated images. The ADAM optimizer \cite{adam_opt} with default settings $\beta_1 = 0.9$ $\beta_2 = 0.999$, and $\epsilon=10^{-8}$ is selected for the optimization. We initialize training parameters of the generator with the pre-trained RRDB \cite{wang2018esrgan} weights and then perform 60k iterations with an initial learning rate of $10^{-4}$ which is halved after 50k iterations. Since wavelet loss is calculated during the training, it does not affect the runtime, hence the inference time of WGSR is the same as the inference time of RRDB \cite{wang2018esrgan}.

\noindent \textbf{Benchmarks and Metrics.} 
To assess the generalization performance of our model, we report results on Set5 \cite{set5_cite}, Set14 \cite{set14_cite}, BSD100 \cite{bsd100_cite}, Urban100 \cite{urban100_cite} and DIV2K \cite{Agustsson_2017_CVPR_Workshops} validation dataset. We report PSNR and SSIM scores on the Y channel to demonstrate the objective quality of generated SR images. The perceptual quality of images is assessed via utilizing the full-reference metrics LPIPS \cite{lpips}, DISTS \cite{dists}, and no-reference metrics NIQE \cite{niqe}, NRQM \cite{Ma_NRQM} and PI \cite{PI} on RGB images for a comprehensive evaluation. We also report LR-PSNR results on benchmarks to verify the LR-consistency of the predicted results. SR predictions must achieve at least 45 dB PSNR between the downsampled version of SR predictions and the corresponding LR images to satisfy the LR-Consistency criterion~\cite{2021_ntire,2022_ntire}.

\begin{figure*}
\centering
\begin{subfigure}{0.16\textwidth}
     \begin{subfigure}{\textwidth}
        \includegraphics[width=\textwidth]{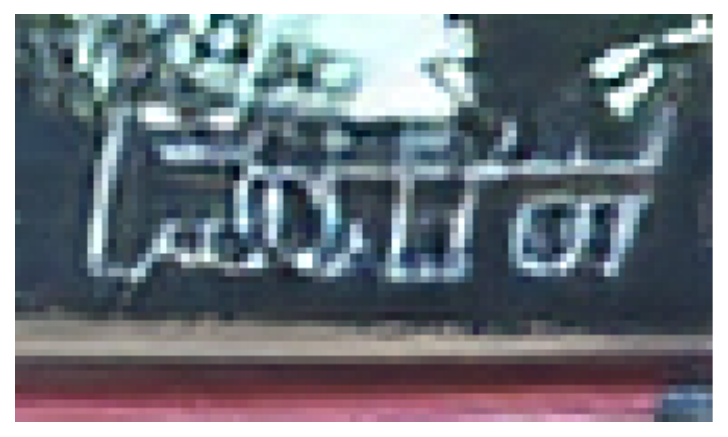} \vspace{-8pt}\\ \scriptsize ESRGAN-FS \cite{freq_sep} \\ (17.98 / 0.191)
    \end{subfigure}
    \begin{subfigure}{\textwidth}
        \includegraphics[width=\textwidth]{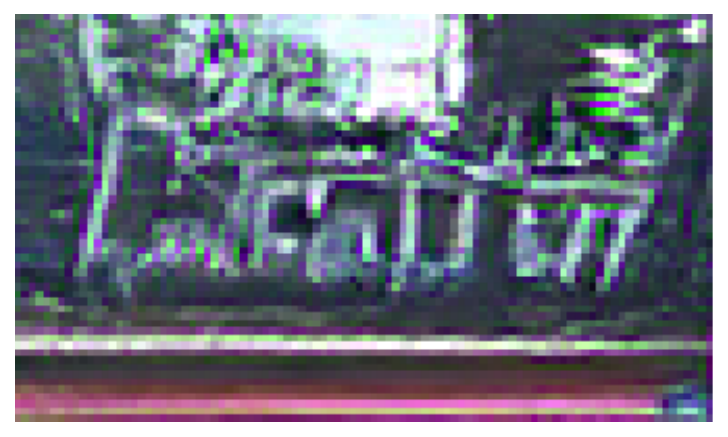} \vspace{-8pt}\\ \scriptsize ESRGAN+ \cite{esrganplus} \\ (17.04 / 0.281)
    \end{subfigure}
    \begin{subfigure}{\textwidth}
        \includegraphics[width=\textwidth]{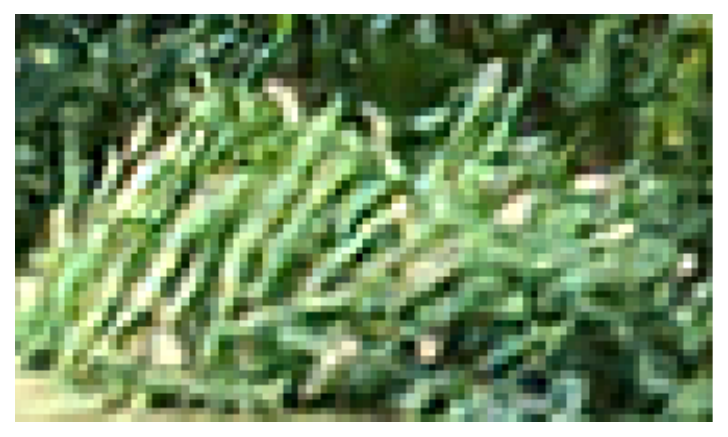} \vspace{-8pt}\\ \scriptsize ESRGAN-FS \cite{freq_sep} \\ (15.07 / 0.273)
    \end{subfigure}
    \begin{subfigure}{\textwidth}
        \includegraphics[width=\textwidth]{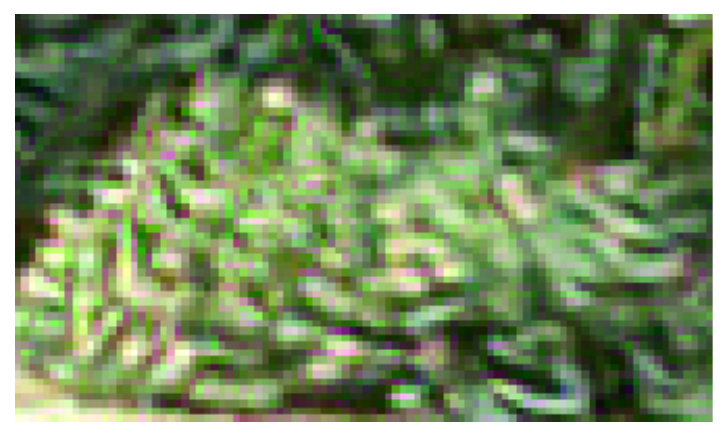} \vspace{-8pt}\\ \scriptsize ESRGAN+ \cite{esrganplus} \\ (16.29 / 0.262)
    \end{subfigure}
\end{subfigure}
\begin{subfigure}{0.16\textwidth}
    \begin{subfigure}{\textwidth}
        \includegraphics[width=\textwidth]{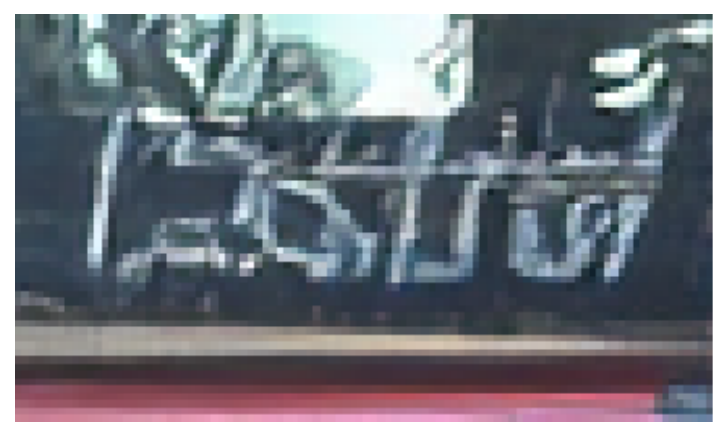} \vspace{-8pt}\\ \scriptsize SPSR \cite{ma_SPSR} \\ (18.71 / 0.196)
    \end{subfigure}
    \begin{subfigure}{\textwidth}
        \includegraphics[width=\textwidth]{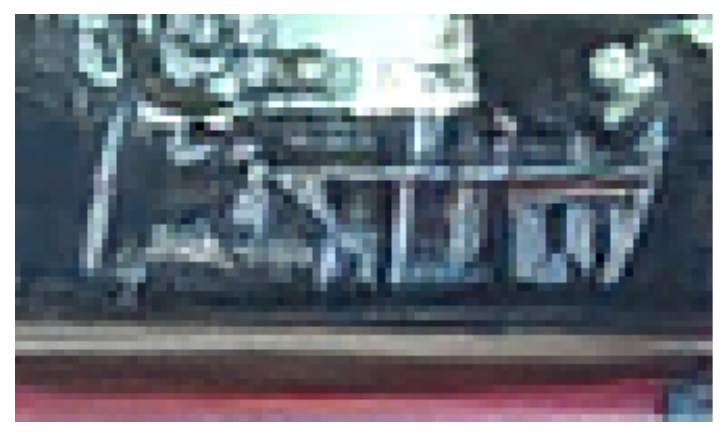} \vspace{-8pt}\\ \scriptsize RankSRGAN \cite{zhang2021ranksrgan} \\ (18.02 / 0.230)
    \end{subfigure}
    \begin{subfigure}{\textwidth}
        \includegraphics[width=\textwidth]{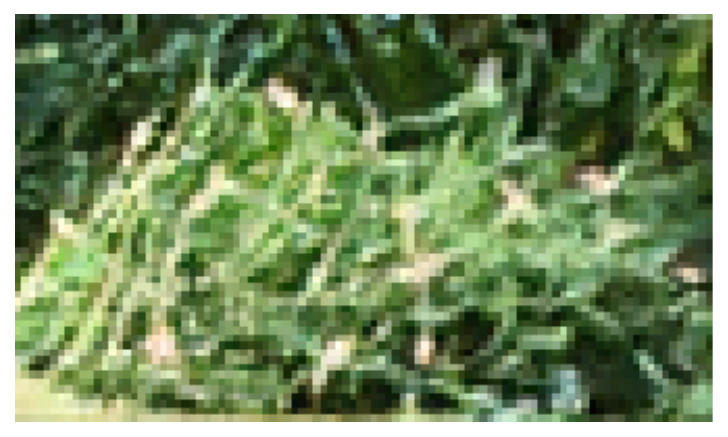} \vspace{-8pt}\\ \scriptsize SPSR \cite{ma_SPSR} \\ (15.70 / 0.248)
    \end{subfigure}
    \begin{subfigure}{\textwidth}
        \includegraphics[width=\textwidth]{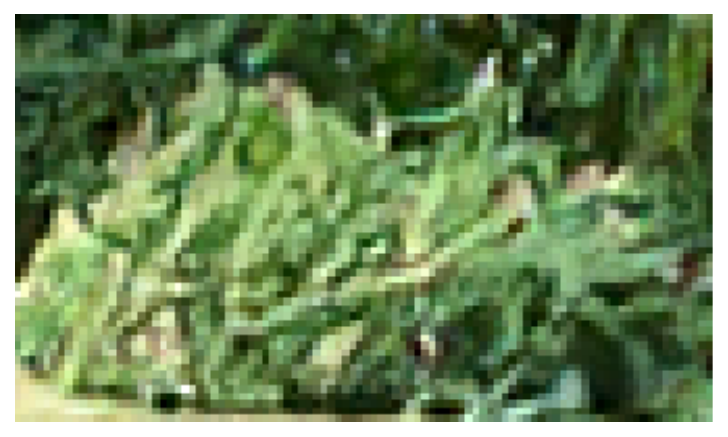} \vspace{-8pt}\\ \scriptsize  RankSRGAN \cite{zhang2021ranksrgan} \\ (16.18 / 0.269)
    \end{subfigure}
\end{subfigure}
\begin{subfigure}{0.16\textwidth}
    \begin{subfigure}{\textwidth}
        \includegraphics[width=\textwidth]{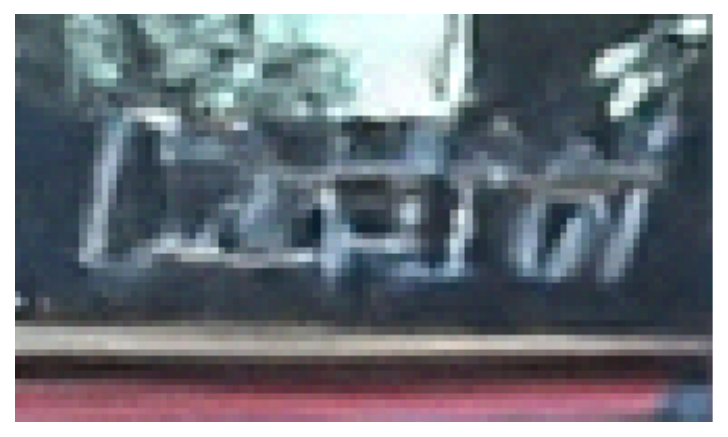} \vspace{-8pt}\\\scriptsize  SRFlow-DA \cite{jo2021srflowda} \\ (20.01 / 0.230)
    \end{subfigure}
    \begin{subfigure}{\textwidth}
        \includegraphics[width=\textwidth]{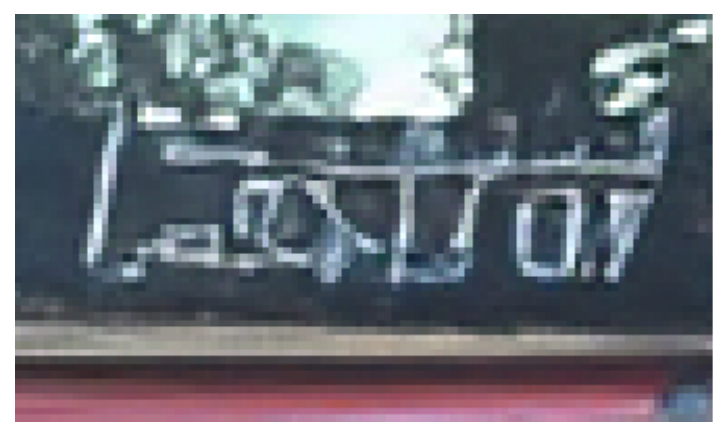} \vspace{-8pt}\\ \scriptsize LDL \cite{details_or_artifacts} \\ (18.50 / 0.190)
    \end{subfigure}
    \begin{subfigure}{\textwidth}
        \includegraphics[width=\textwidth]{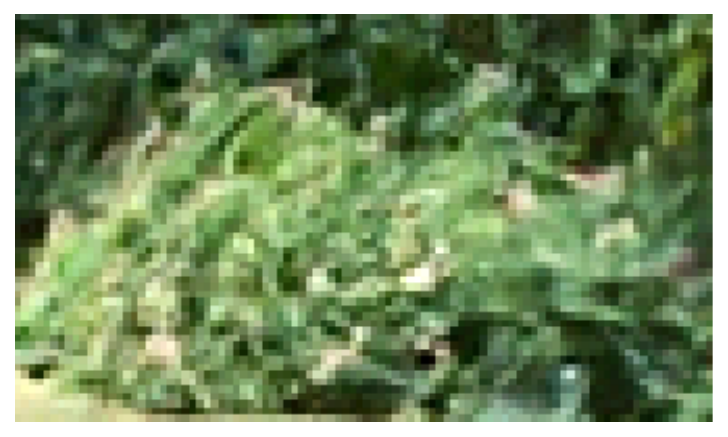} \vspace{-8pt}\\ \scriptsize SRFlow-DA \cite{jo2021srflowda} \\ (16.33 / 0.224)
    \end{subfigure}
    \begin{subfigure}{\textwidth}
        \includegraphics[width=\textwidth]{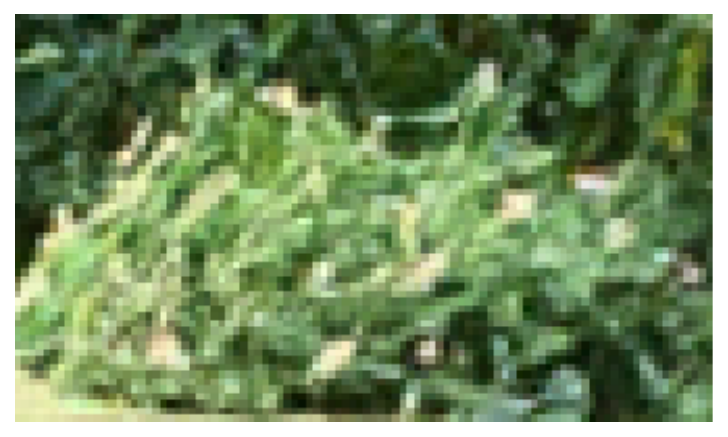} \vspace{-8pt}\\ \scriptsize LDL \cite{details_or_artifacts} \\ (16.78 / 0.240)
    \end{subfigure}
\end{subfigure}
\begin{subfigure}{0.16\textwidth}
    \begin{subfigure}{\textwidth}
        \includegraphics[width=\textwidth]{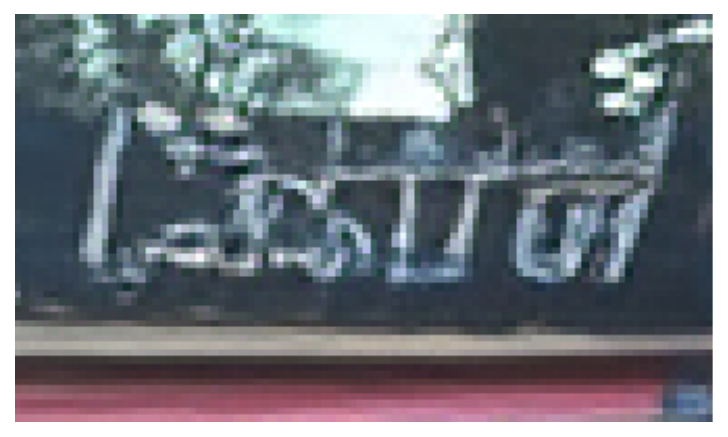} \vspace{-8pt}\\ \scriptsize FxSR \cite{fxsr} \\ (19.20 / 0.212)
    \end{subfigure}
            \begin{subfigure}{\textwidth}
        \includegraphics[width=\textwidth]{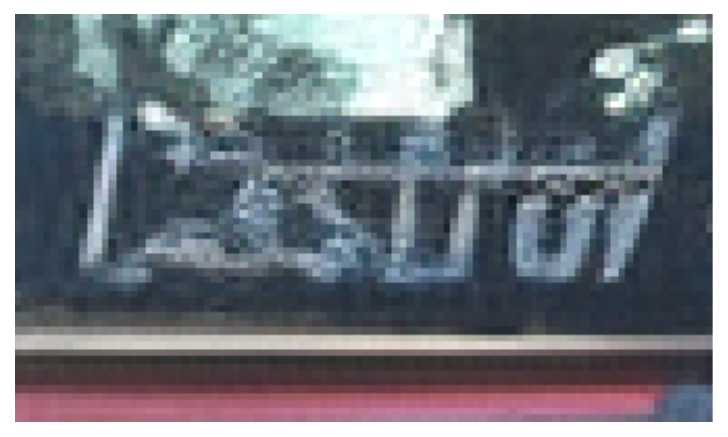} \vspace{-8pt}\\ \scriptsize PDASR \cite{PDASR} \\ (19.71 / 0.237)
    \end{subfigure}
    \begin{subfigure}{\textwidth}
        \includegraphics[width=\textwidth]{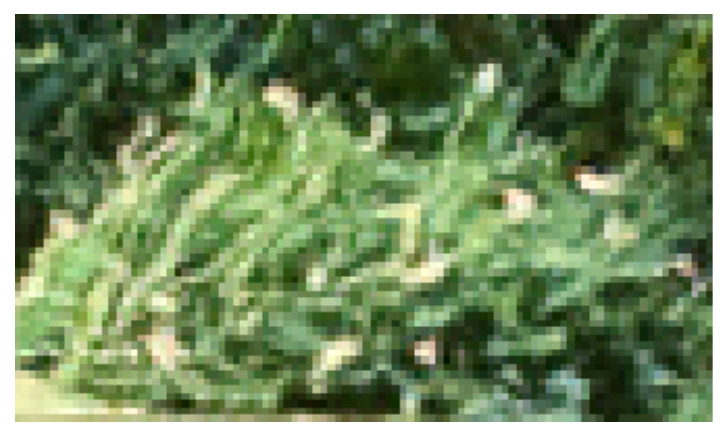} \vspace{-8pt}\\ \scriptsize FxSR \cite{fxsr} \\ (16.51 / 0.239)
    \end{subfigure}
    \begin{subfigure}{\textwidth}
        \includegraphics[width=\textwidth]{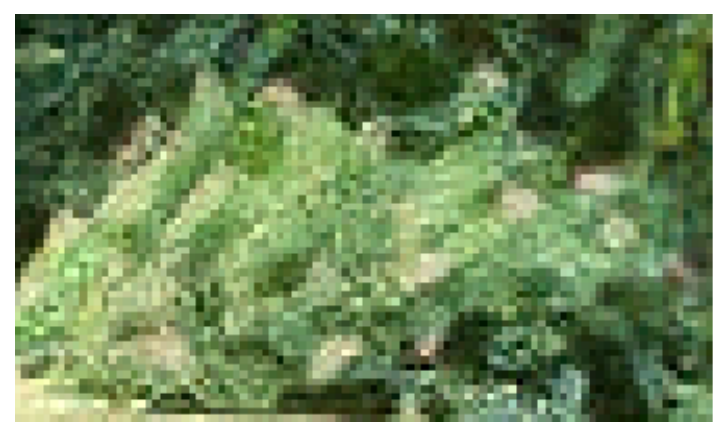} \vspace{-8pt}\\ \scriptsize PDASR \cite{PDASR} \\ (16.92 / 0.201)
    \end{subfigure}
\end{subfigure}
\begin{subfigure}{0.16\textwidth}
    \begin{subfigure}{\textwidth}
        \includegraphics[width=\textwidth]{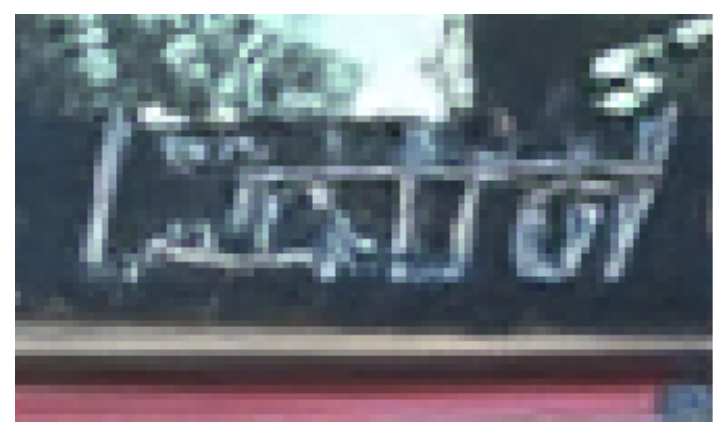} \vspace{-8pt}\\ \scriptsize SROOE \cite{srooe_Park_2023_CVPR} \\ (19.93 / 0.206)
    \end{subfigure}
    \begin{subfigure}{\textwidth}
        \includegraphics[width=\textwidth]{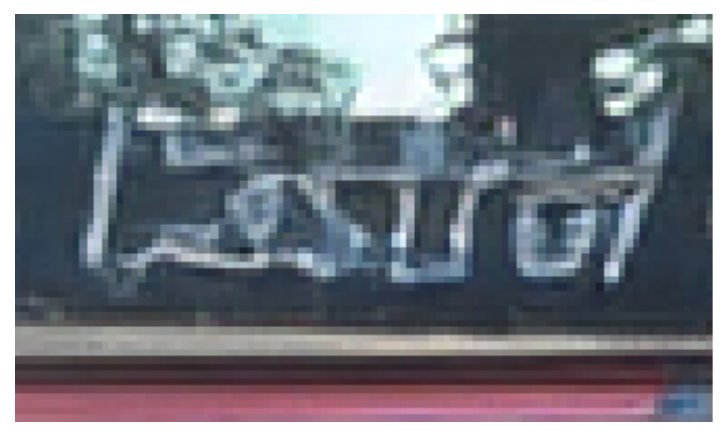} \vspace{-8pt}\\ \scriptsize  DualFormer \cite{dualformer_luo2023effectiveness} \\ (19.62 / 0.213)
    \end{subfigure}
    \begin{subfigure}{\textwidth}
        \includegraphics[width=\textwidth]{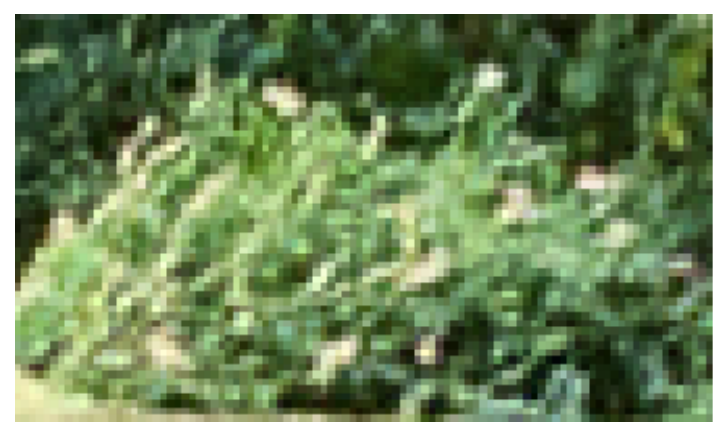} \vspace{-8pt}\\ \scriptsize SROOE \cite{srooe_Park_2023_CVPR} \\ (16.50 / 0.241)
    \end{subfigure}
    \begin{subfigure}{\textwidth}
        \includegraphics[width=\textwidth]{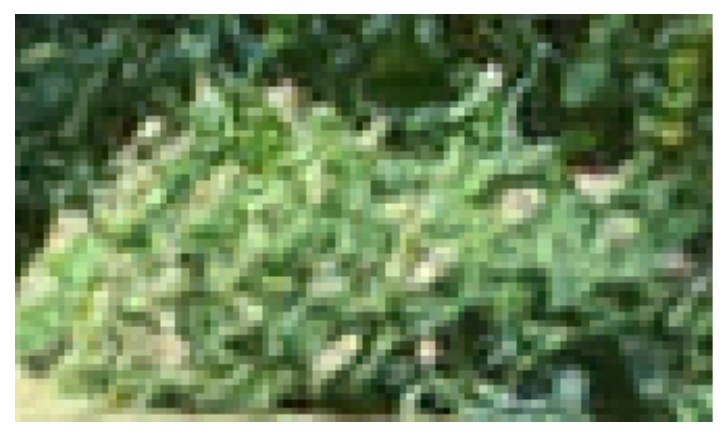} \vspace{-8pt}\\ \scriptsize  DualFormer \cite{dualformer_luo2023effectiveness} \\ (16.84 / 0.230)
    \end{subfigure}
\end{subfigure}
\begin{subfigure}{0.16\textwidth}
    \begin{subfigure}{\textwidth}
        \includegraphics[width=\textwidth]{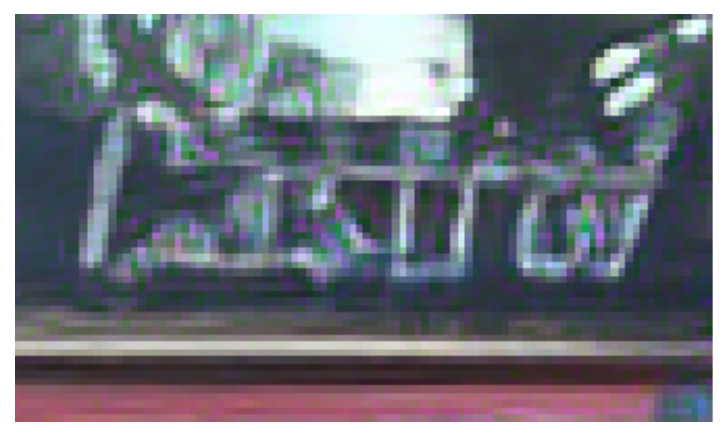} \vspace{-8pt}\\ \scriptsize WGSR (2-lvl) \\(19.92 / 0.250)
    \end{subfigure}
        \begin{subfigure}{\textwidth}
        \includegraphics[width=\textwidth]{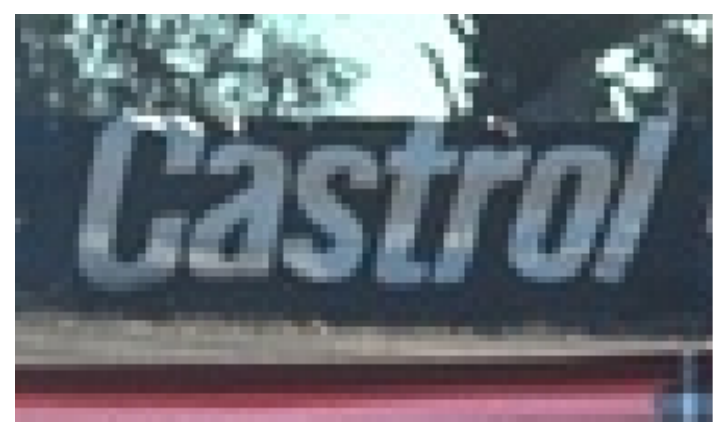} \vspace{-8pt}\\ \scriptsize HR (img-41)\\ (PSNR / DISTS\cite{dists})
    \end{subfigure}
    \begin{subfigure}{\textwidth}
        \includegraphics[width=\textwidth]{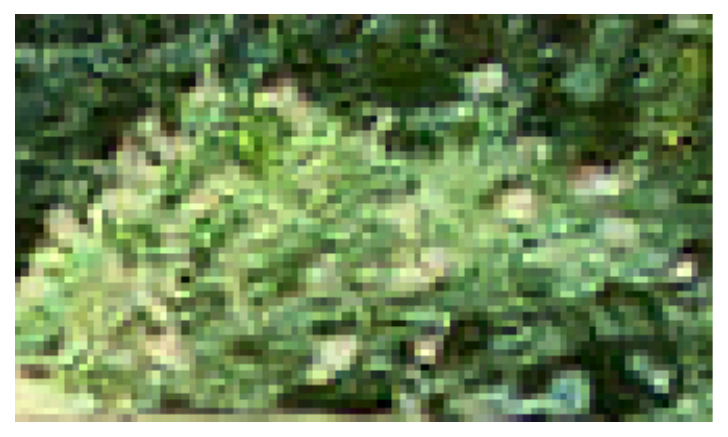} \vspace{-8pt}\\ \scriptsize  WGSR (2-lvl) \\ (17.26 / 0.196)
    \end{subfigure}
    \begin{subfigure}{\textwidth}
        \includegraphics[width=\textwidth]{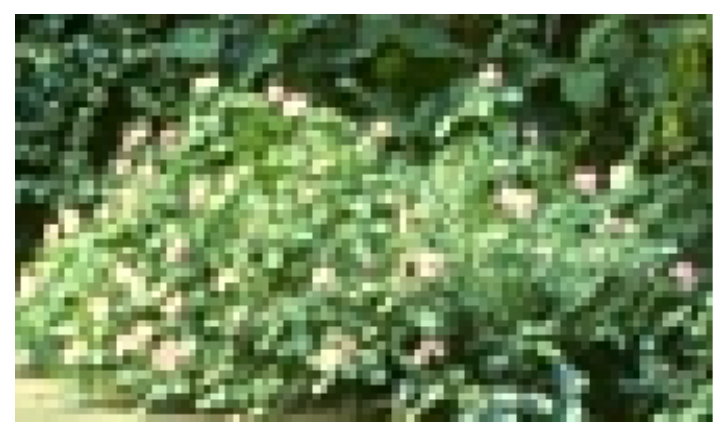} \vspace{-8pt}\\ \scriptsize HR (img-21)\\ (PSNR / DISTS\cite{dists})
    \end{subfigure}
\end{subfigure} \vspace{-3pt}
\caption{Visual comparison of the proposed wavelet-guided perceptual optimization method with the state-of-the-art methods for $\times$4 SR on natural images from BSD100 validation set. Our method WGSR with 2-level SWT provides the best balance between perception-distortion trade-off for natural images and it has clear advantages in reconstructing realistic HF details while inhibiting artifacts. Additional visual comparisons can be found in the supplementary materials.}
\label{fig:qual_results} 
\end{figure*}

\subsection{Comparison with the state-of-the-art}
\noindent
\textbf{Quantitative Comparison.}
Table \ref{table:quantitative_results} demonstrates quantitative comparison for $\times$4 SR methods and our proposed approach WGSR.  We compare our method with the existing state-of-the-art methods including ESRGAN-FS \cite{freq_sep}, ESRGAN+ \cite{esrganplus}, SPSR \cite{ma_SPSR}, RankSRGAN \cite{zhang2021ranksrgan}, SRFlow-DA (heat=0.9) \cite{jo2021srflowda}, LDL \cite{details_or_artifacts},  FxSR (t=0.8) \cite{fxsr}, PDASR \cite{PDASR} and SROOE (t=0.9)\cite{srooe_Park_2023_CVPR}. Our method WGSR improves the perceptual quality and the reconstruction accuracy simultaneously. Specifically, the table shows that our method yields the best perceptual scores in terms of NIQE, NRQM, and PI without significantly compromising objective quality. Also, in terms of distortion-oriented metrics such as PSNR and SSIM, our method provides better fidelity scores compared to other GAN-SR approaches. Our proposed network WGSR also exceeds 45 dB LR-PSNR which guarantees the original information conveyed by the LR images is preserved, thus, it does not suffer from LR-consistency, unlike ESRGAN-FS \cite{freq_sep}, ESRGAN+ \cite{esrganplus}, SPSR \cite{ma_SPSR} and RankSRGAN \cite{zhang2021ranksrgan}. To conclude, our proposed wavelet guidance for the optimization objective preserves the LR manifold and generates photo-realistic high perceptual quality SR images. 

\noindent
\textbf{Qualitative Comparison.}
Visual comparisons among $\times$4 SR approaches and WGSR are presented in Fig.~\ref{fig:first_img} and~\ref{fig:qual_results}. Similar conclusions to the quantitative comparisons can be drawn from qualitative comparisons. We observe that all GAN-SR results including ESRGAN-FS \cite{freq_sep}, ESRGAN+ \cite{esrganplus}, SPSR \cite{ma_SPSR}, RankSRGAN \cite{zhang2021ranksrgan}, LDL \cite{details_or_artifacts} and FxSR \cite{fxsr} produce visible artifacts and experience excessive sharpness. On the other hand, our method WGSR is able to reconstruct the genuine image details with high reconstruction accuracy including the regions with regular patterns and the areas containing fine details such as the light pink flowers on the bush (Fig. \ref{fig:qual_results}). Moreover, the visual results presented in Fig. \ref{fig:first_img} certainly demonstrate the reconstruction power of WGSR when it comes to information-centric applications. The other state-of-the-art GAN-SR methods cannot recover the correct number ``45". In contrast, WGSR is the clear winner for that image patch by learning to control artifacts while providing genuine image details. These improvements show that the wavelet-domain losses is a suitable optimization objective to train GAN-SR models to obtain photo-realistic, high-quality and accurate SR images. 

\begin{figure}
\centering
\begin{subfigure}{0.15\textwidth} 
     \begin{subfigure}{\textwidth}
        \includegraphics[width=\textwidth]{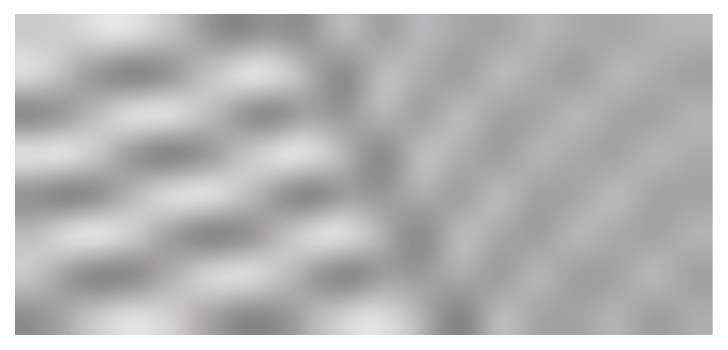} \vspace{-16pt}
    \end{subfigure} 
    \caption*{Bicubic interp}
     \begin{subfigure}{\textwidth}
        \includegraphics[width=\textwidth]{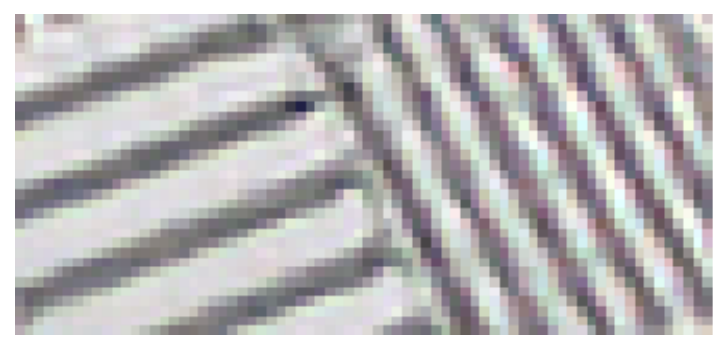} \vspace{-16pt}
    \end{subfigure} 
    \caption*{WGSR (1-lvl)}
\end{subfigure}
\begin{subfigure}{0.15\textwidth}
     \begin{subfigure}{\textwidth}
        \includegraphics[width=\textwidth]{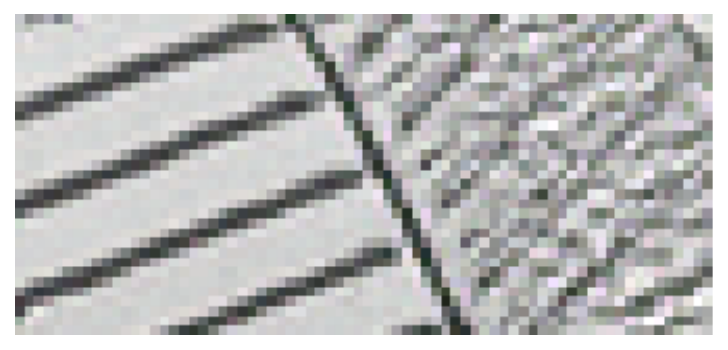} \vspace{-16pt}
    \end{subfigure} 
    \caption*{FxSR \cite{fxsr}}
     \begin{subfigure}{\textwidth}
        \includegraphics[width=\textwidth]{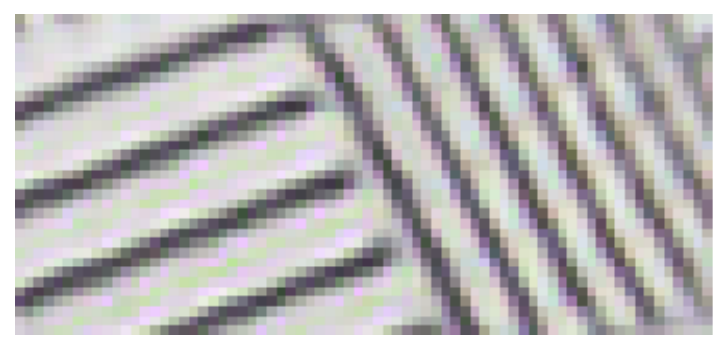} \vspace{-16pt}
    \end{subfigure} 
    \caption*{WGSR (2-lvl)}
\end{subfigure}
\begin{subfigure}{0.15\textwidth}
         \begin{subfigure}{\textwidth}
        \includegraphics[width=\textwidth]{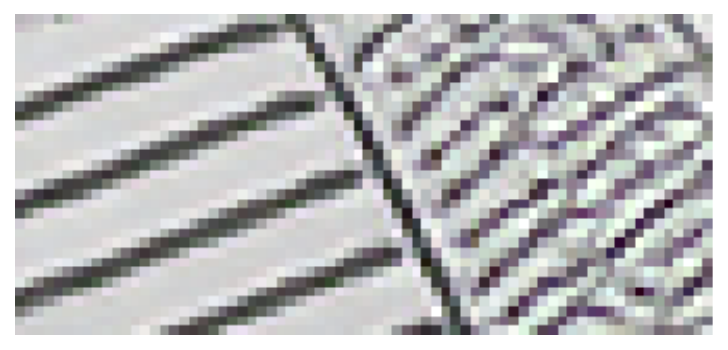} \vspace{-16pt}
    \end{subfigure} 
    \caption*{SROOE \cite{srooe_Park_2023_CVPR}}
     \begin{subfigure}{\textwidth}
        \includegraphics[width=\textwidth]{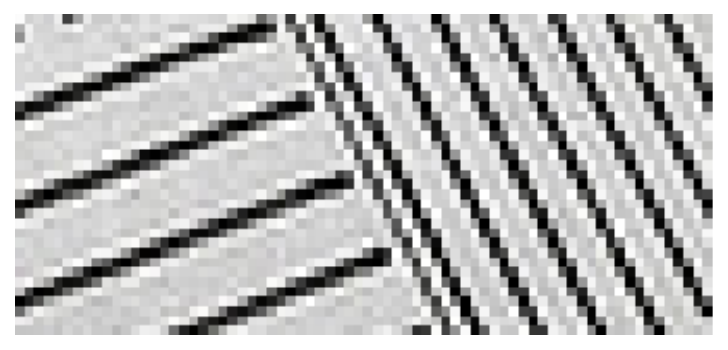} \vspace{-16pt}
    \end{subfigure} 
    \caption*{HR ground-truth}
\end{subfigure} \vspace{-6pt}
\caption{Visual comparison of different models on an image from Urban100 (img-92) dataset. FxSR \cite{fxsr}, SROOE \cite{srooe_Park_2023_CVPR} and WGSR with 1-level SWT show aliasing~artifacts whereas WGSR with 2-level SWT recovers all structures at different scales.}
\label{fig:wavelet_level} 
\end{figure}

\noindent
\textbf{SWT Decomposition Levels.}
The number of levels of the~SWT decomposition is another parameter that offers flexibility on controlling genuine details vs. artifacts and affects the SR performance. The best number of levels depends on the scale and orientation of structures appearing in LR images. An example image crop containing lines with different orientation and spatial frequencies is shown in Fig.~\ref{fig:wavelet_level}. The state-of-the-art GAN-based SR methods FxSR~\cite{fxsr} and SROOE \cite{srooe_Park_2023_CVPR} are unable to recover the~correct structure on the right section of the crop. Here, our WGSR model with 1-level SWT can recover the correct orientiation of lines but some visible aliasing remains. The~best visual result can be obtained when 2-level SWT is used by further decomposition of the LL subband of 1-level SWT into 4 subbands (L-LL, L-LH, L-HL, L-HH) and keeping the details subbands as in 1-level SWT. That~is, after applying 2-level SWT, we obtain 7 subbands which brings additional flexibility in weighting loss terms for each subband. In our results, the weight parameters for 2-level decomposition is set to $\lambda_{L-LL}$ = 0.1, $\lambda_{L-LH}$ = $\lambda_{L-HL}$ = 0.01 and $\lambda_{L-HH}$ = 0.05, $\lambda_{LH}$ = $\lambda_{HL}$ = 0.1 and $\lambda_{HH}$ = 0.05. Note that the detail (LH, HL, and HH) subbands of the 2-level decomposition are the same as in 1-level decomposition and they are assigned the same weights. We observe that computing losses on 2-level SWT manages to recover genuine details and structures when its level-2 (mid) HF subbands are penalized more in fidelity losses. 

\noindent
\textbf{The Choice of Wavelet Family.}
In order to investigate the~effect of the choice of wavelet family on our results, we~conduct experiments with a large selection of wavelet filters including haar; db7 and db19 from Daubechies; sym7 and sym19 from Symlets; bior2.6 and bior4.4 from Biorthogonal wavelet families. 

\begin{figure}
\begin{subfigure}{0.45\textwidth}
    \begin{subfigure}{\textwidth}
        \includegraphics[width=\textwidth]{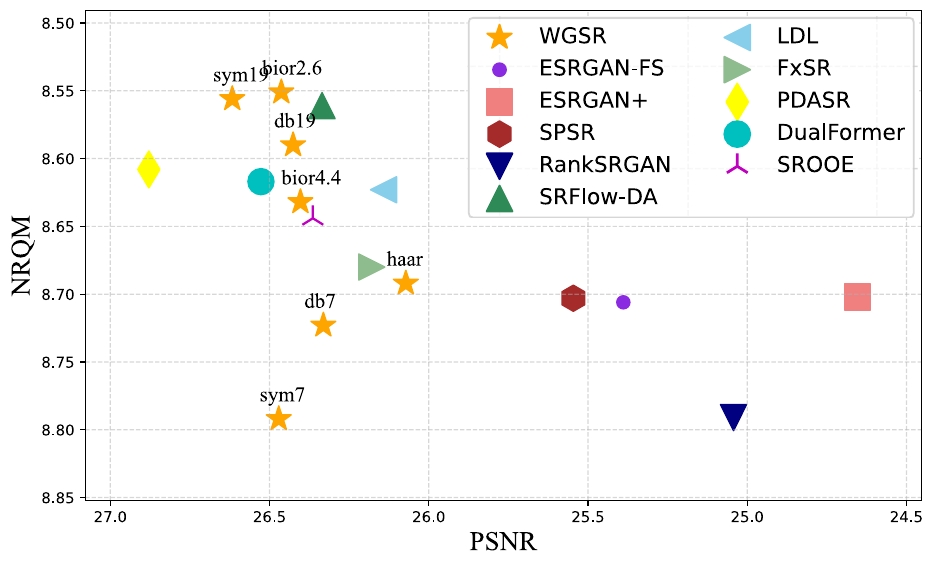} \vspace{-16pt}
    \end{subfigure}
\end{subfigure} \vspace{-6pt}
\caption{Perception-distortion trade-off performance of our method WGSR with different wavelet families indicated as orange stars and comparison of other state-of-the-art methods on the PSNR-NRQM plane for BSD100 \cite{bsd100_cite} dataset.}
\label{fig:bsd100_wavelet} 
\end{figure}

The PD trade-off performance of our WGSR model with different wavelet families on BSD100~\cite{bsd100_cite} benchmark is shown in Fig.~\ref{fig:bsd100_wavelet}. We observe that the PD trade-off performance varies according to the choice of wavelet family. The~best objective quality is provided by the Symlet ``sym19" filter and the best perceptual quality among all solutions is achieved by the Daubechies ``db7" filter. The~results show that the best trade-off point is achieved by the~Symlet ``sym7" filter, since it is the closest to the lower left corner of the PSNR-NRQM plane. Hence, we utilize `sym7" wavelet filter in our results.

\subsection{Ablation Study}
We conduct ablation studies to investigate the effect of each loss term in our WGSR method including the fidelity $l_1$, adversarial $L_{adv,G}$ and the perceptual loss $L_{perc}$ in eqn.~\ref{eq:loss_adv_gan}. Results are reported in Table \ref{tab:ablation_table}. $\#$0 refers to the baseline ESRGAN \cite{wang2018esrgan}, where $l_1$ and $L_{adv,G}$ are computed in the~RGB domain, and  $L_{perc}$ is taken as LPIPS. 

In $\#$1, we change $L_{perc}$ from LPIPS~\cite{lpips} to DISTS \cite{dists}, which results in an increase of 0.3 dB and 1.6$\%$ in objective and perceptual performance, respectively, compared to baseline $\#$0.  Similar improvements can also be observed from $\#$4 to $\#$5, where both $l_1$ and $L_{adv,G}$ are computed in the SWT domain. These results validate using DISTS  instead of LPIPS navigates the SR model to a better PD point since both objective and perceptual performance improves.

Next, we investigate the effect of computing $l_1$ and $L_{adv,G}$ losses in the RGB vs. SWT domain. 
In $\#$2, the~$l_1$ fidelity loss is calculated in the SWT domain. We observe that the objective quality is improved by almost 1~dB without any change in perceptual quality. This clearly demonstrates the generated details can be better controlled by enforcing fidelity in the SWT subbands as opposed to RGB images. On the other hand, calculation of $L_{adv}$ in the SWT domain in $\#$3 favors the perceptual quality. Finally, combining all SWT domain losses in $\#$5 (WGSR) achieves the~best trade-off between objective and perceptual quality.

\begin{table}
\caption{Comparison of fidelity and perceptual performance according to the selection of domain of $l_1$ and $L_{adv,G}$ and type of $L_{perc}$ evaluated on BSD100 \cite{bsd100_cite} benchmark. The best and the~second-best are marked in \textbf{bold} and \underline{underlined}, respectively. We see that the proposed combination of losses ($\#$5) achieves near 2 dB PSNR gain and better PI score compared to baseline ($\#$0).} \vspace{-6pt}
    \centering
    \begin{tabular}{cccccc}
    \# & $l_1$ & $L_{adv,G}$ & $L_{perc}$ &PSNR $\uparrow$ & PI $\downarrow$ \\ \hline
    0 & RGB & RGB & LPIPS & 24.506 & 2.543 \\
    1 & RGB & RGB & DISTS & 24.812 & 2.502 \\
    2 & SWT & RGB & DISTS &  25.622 & 2.501 \\
    3 & RGB & SWT & DISTS & 24.746 & \textbf{2.443} \\
    4 & SWT & SWT & LPIPS & \underline{25.859} & 2.466 \\
    5 & SWT & SWT & DISTS & \textbf{26.331} & \underline{2.453} \\
    \end{tabular}
    \label{tab:ablation_table}
\end{table}

\subsection{Discussion of Limitations}
Although the proposed training method is effective in improving both fidelity and perceptual quality of SR images, there are some challenges that remain: 

i) Neither the PSNR nor any quantitative perceptual scores are good indicators of visual artifacts. We demonstrate that WGSR is effective in suppressing visual artifacts in Figures \ref{fig:first_img}, \ref{fig:qual_results}, and \ref{fig:wavelet_level}. However, this visual performance does not reflect on quantitative measures.

ii) Determining the best selection of weights on different SWT-domain loss terms is difficult. During our search for the best weights, we noticed decreasing weights on fidelity losses for LH and HL subbands causes fidelity scores to drop, and increasing the weight of the fidelity term on the HH subband decreases perceptual quality. On the other hand, higher $\lambda_{adv}$ or $\lambda_{perc}$ lead to improvement of perceptual quality at the expense of fidelity. Hence, the selection of weights leads to different PD trade-off points. In summary, while our results demonstrate training by wavelet-domain losses steers towards a better PD point, we believe there is still room for further improvements in discriminating genuine image details from artifacts.
\section{Conclusion}

The PD trade-off hypothesis states the impossibility of improving both fidelity and perceptual quality simultaneously beyond a theoretical limit, which is unknown in practical settings. This paper shows we can improve both fidelity~(PSNR) and perceptual quality (NRQM) compared to most of the SOTA methods, while we can only improve on one with just a small compromise on the other vs. some other methods. Hence, we claim we can reach a better PD trade-off with the guidance of wavelet-domain losses. In~particular, we propose a novel GAN-based SR model training method, which utilizes a weighted combinations of wavelet-domain losses. By controlling the strength of fidelity and adversarial losses according to the scale and orientation of image features in different subbands, our model is capable of learning genuine image details with high reconstruction accuracy without suffering HF artifacts and hallucinations.  Extensive experiments on widely used benchmark datasets demonstrate that WGSR outperforms existing GAN-SR methods quantitatively and qualitatively, and provides better PD trade-off performance. The proposed method for adversarial training is generic in the sense that any off-the-shelf GAN-SR model can be easily plugged into this framework to benefit from wavelet guidance.

\vspace{10pt}



\newpage

{
    \small
    \bibliographystyle{ieeenat_fullname}
    \bibliography{source}

\begin{thebibliography}{72}
\providecommand{\natexlab}[1]{#1}
\providecommand{\url}[1]{\texttt{#1}}
\expandafter\ifx\csname urlstyle\endcsname\relax
  \providecommand{\doi}[1]{doi: #1}\else
  \providecommand{\doi}{doi: \begingroup \urlstyle{rm}\Url}\fi

\bibitem[Agustsson and Timofte(2017)]{Agustsson_2017_CVPR_Workshops}
E Agustsson and R. Timofte.
\newblock {NTIRE 2017 Challenge} on single image super-resolution: Dataset and study.
\newblock In \emph{IEEE/CVF Conf. on Comp. Vision and Patt. Recog. (CVPR) Workshops}, 2017.

\bibitem[Bevilacqua et~al.(2012)Bevilacqua, Roumy, Guillemot, and line Alberi~Morel]{set5_cite}
Marco Bevilacqua, Aline Roumy, Christine Guillemot, and Marie line Alberi~Morel.
\newblock Low-complexity single-image super-resolution based on nonnegative neighbor embedding.
\newblock In \emph{Proc. of the British Machine Vision Conference}, pages 135.1--135.10, 2012.

\bibitem[Blau and Michaeli(2018)]{Blau_2018}
Yochai Blau and Tomer Michaeli.
\newblock The perception-distortion tradeoff.
\newblock \emph{IEEE/CVF Conf. on Computer Vision and Pattern Recognition}, 2018.

\bibitem[Blau et~al.(2018)Blau, Mechrez, Timofte, Michaeli, and Zelnik-Manor]{PI}
Yochai Blau, Roey Mechrez, Radu Timofte, Tomer Michaeli, and Lihi Zelnik-Manor.
\newblock 2018 pirm challenge on perceptual image super-resolution.
\newblock In \emph{European Conf. Comp. Vision (ECCV) Workshops}, 2018.

\bibitem[Chandaliya and Nain(2022)]{awgan_2022}
Praveen~Kumar Chandaliya and Neeta Nain.
\newblock Aw-gan: Face aging and rejuvenation using attention with wavelet gan.
\newblock \emph{Neural Comput. Appl.}, 35\penalty0 (3):\penalty0 2811–2825, 2022.

\bibitem[Chen et~al.(2021)Chen, Li, Jin, Liu, and Li]{ssd_gan_chen2021ssd}
Yuanqi Chen, Ge Li, Cece Jin, Shan Liu, and Thomas Li.
\newblock Ssd-gan: measuring the realness in the spatial and spectral domains.
\newblock In \emph{Proceedings of the AAAI Conference on Artificial Intelligence}, pages 1105--1112, 2021.

\bibitem[Dai et~al.(2019)Dai, Cai, Zhang, Xia, and Zhang]{dai_attention}
Tao Dai, Jianrui Cai, Yongbing Zhang, Shu-Tao Xia, and Lei Zhang.
\newblock Second-order attention network for single image super-resolution.
\newblock In \emph{IEEE/CVF Conf. on Comp. Vision and Pattern Recog. (CVPR)}, pages 11057--11066, 2019.

\bibitem[Deng(2018)]{first_deng}
Xin Deng.
\newblock Enhancing image quality via style transfer for single image super-resolution.
\newblock \emph{IEEE Signal Processing Letters}, 25\penalty0 (4):\penalty0 571--575, 2018.

\bibitem[Deng et~al.(2019)Deng, Yang, Xu, and Dragotti]{Deng2019WaveletDS}
Xin Deng, Ren Yang, Mai Xu, and Pier~Luigi Dragotti.
\newblock Wavelet domain style transfer for an effective perception-distortion tradeoff in single image super-resolution.
\newblock \emph{IEEE/CVF Int. Conf. on Computer Vision (ICCV)}, pages 3076--3085, 2019.

\bibitem[Ding et~al.(2020)Ding, Ma, Wang, and Simoncelli]{dists}
K. Ding, K. Ma, S. Wang, and E.~P. Simoncelli.
\newblock Image quality assessment: Unifying structure and texture similarity.
\newblock \emph{IEEE Trans. on Patt Anal. and Mach. Intel.}, 44:\penalty0 2567--2581, 2020.

\bibitem[Dong et~al.(2014)Dong, Loy, He, and Tang]{dong_srcnn}
Chao Dong, Chen~Change Loy, Kaiming He, and Xiaoou Tang.
\newblock Learning a deep convolutional network for image super-resolution.
\newblock In \emph{European Conf. Comp. Vision (ECCV)}, pages 184--199, 2014.

\bibitem[Fritsche et~al.(2019)Fritsche, Gu, and Timofte]{freq_sep}
Manuel Fritsche, Shuhang Gu, and Radu Timofte.
\newblock Frequency separation for real-world super-resolution.
\newblock In \emph{IEEE/CVF Int. Conf. on Computer Vision Workshop (ICCVW)}, pages 3599--3608, 2019.

\bibitem[Fuoli et~al.(2021)Fuoli, Gool, and Timofte]{Fuoli2021FourierSL}
Dario Fuoli, Luc~Van Gool, and Radu Timofte.
\newblock Fourier space losses for efficient perceptual image super-resolution.
\newblock \emph{IEEE/CVF Int. Conf. on Computer Vision (ICCV)}, pages 2340--2349, 2021.

\bibitem[Gal et~al.(2021)Gal, Hochberg, Bermano, and Cohen-Or]{gal2021swagan}
Rinon Gal, Dana~Cohen Hochberg, Amit Bermano, and Daniel Cohen-Or.
\newblock Swagan: A style-based wavelet-driven generative model.
\newblock \emph{ACM Transactions on Graphics (TOG)}, 40\penalty0 (4):\penalty0 1--11, 2021.

\bibitem[Gao et~al.(2023)Gao, Liu, Zeng, Xu, Li, Luo, Liu, Zhen, and Zhang]{gao2023implicit}
Sicheng Gao, Xuhui Liu, Bohan Zeng, Sheng Xu, Yanjing Li, Xiaoyan Luo, Jianzhuang Liu, Xiantong Zhen, and Baochang Zhang.
\newblock Implicit diffusion models for continuous super-resolution.
\newblock In \emph{IEEE/CVF Conf. on Comp. Vision and Patt. Recog.}, pages 10021--10030, 2023.

\bibitem[Goodfellow et~al.(2014)Goodfellow, Pouget-Abadie, Mirza, Xu, Warde-Farley, Ozair, Courville, and Bengio]{ian_gan}
Ian Goodfellow, Jean Pouget-Abadie, Mehdi Mirza, Bing Xu, David Warde-Farley, Sherjil Ozair, Aaron Courville, and Yoshua Bengio.
\newblock Generative adversarial nets.
\newblock In \emph{Advances in Neural Information Processing Systems}, 2014.

\bibitem[Gu et~al.(2022)Gu, Cai, Dong, et~al.]{Gu_2022_CVPR}
Jinjin Gu, Haoming Cai, Chao Dong, et~al.
\newblock Ntire 2022 challenge on perceptual image quality assessment.
\newblock In \emph{IEEE/CVF Conf. on Computer Vision and Pattern Recognition (CVPR) Workshops}, pages 951--967, 2022.

\bibitem[Guo et~al.(2017)Guo, Mousavi, Vu, and Monga]{DWSR_guo2017deep}
T. Guo, H.~S. Mousavi, T.~H. Vu, and V. Monga.
\newblock Deep wavelet prediction for image super-resolution.
\newblock In \emph{IEEE/CVF \hspace{-3pt}Conf. \hspace{-3pt}Comp.\hspace{-3pt} Vis. \hspace{-2pt}and\hspace{-1pt} Patt. \hspace{-3pt}Recog.\hspace{-3pt} (CVPRW)}, pages 104--113, 2017.

\bibitem[Huang et~al.(2017)Huang, He, Sun, and Tan]{wavelet_srnet_huang2017wavelet}
H. Huang, R. He, Z. Sun, and T. Tan.
\newblock Wavelet-{SRnet}: {A}~wavelet-based {CNN} for multi-scale face super resolution.
\newblock In \emph{IEEE Int. Conf. Comp. Vis. \hspace{-2pt}(ICCV)}, pages 1689--1697, 2017.

\bibitem[Huang et~al.(2015)Huang, Singh, and Ahuja]{urban100_cite}
Jia-Bin Huang, Abhishek Singh, and Narendra Ahuja.
\newblock Single image super-resolution from transformed self-exemplars.
\newblock In \emph{IEEE Conf. on Comp. Vision and Patt. Recog. (CVPR)}, pages 5197--5206, 2015.

\bibitem[Jawerth and Sweldens(1994)]{wavelet_doc}
Bjorn Jawerth and Wim Sweldens.
\newblock An overview of wavelet based multiresolution analyses.
\newblock \emph{SIAM Review}, 36\penalty0 (3):\penalty0 377--412, 1994.

\bibitem[Jiang et~al.(2021)Jiang, Dai, Wu, and Loy]{jiang2021focal}
Liming Jiang, Bo Dai, Wayne Wu, and Chen~Change Loy.
\newblock Focal frequency loss for image reconstruction and synthesis.
\newblock In \emph{Int. Conf. Comp. Vision (ICCV)}, 2021.

\bibitem[Jo et~al.(2021)Jo, Yang, and Kim]{jo2021srflowda}
Younghyun Jo, Sejong Yang, and Seon~Joo Kim.
\newblock Srflow-da: Super-resolution using normalizing flow with deep convolutional block.
\newblock In \emph{IEEE/CVF Conf. on Computer Vision and Pattern Recognition (CVPR) Workshops}, 2021.

\bibitem[Johnson et~al.(2016)Johnson, Alahi, and Fei-Fei]{johnson2016perceptual}
Justin Johnson, Alexandre Alahi, and Li Fei-Fei.
\newblock Perceptual losses for real-time style transfer and super-resolution.
\newblock In \emph{Computer Vision--ECCV 2016: 14th European Conference, Amsterdam, The Netherlands, October 11-14, 2016, Proceedings, Part II 14}, pages 694--711. Springer, 2016.

\bibitem[Jolicoeur-Martineau(2018)]{jolicoeur2018relativistic}
Alexia Jolicoeur-Martineau.
\newblock The relativistic discriminator: a key element missing from standard gan.
\newblock \emph{arXiv preprint arXiv:1807.00734}, 2018.

\bibitem[Kim et~al.(2016)Kim, Lee, and Lee]{kim2016accurate}
J. Kim, J.~Kwon Lee, and K.~Mu Lee.
\newblock Accurate image super-resolution using very deep convolutional networks.
\newblock \emph{IEEE/CVF Conf. on Comp. Vision and Patt. Recog. (CVPR)}, pages 1646--1654, 2016.

\bibitem[Kim and Son(2021)]{ncsr}
Y. Kim and D. Son.
\newblock Noise conditional flow model for learning the super-resolution space.
\newblock In \emph{IEEE/CVF Conf. on Comp. Vision and Patt. Recog. (CVPR)}, pages 424--432, 2021.

\bibitem[Kingma and Ba(2015)]{adam_opt}
Diederik~P. Kingma and Jimmy Ba.
\newblock Adam: {A} method for stochastic optimization.
\newblock In \emph{Int. Conf. on Learning Representations, {ICLR} San Diego, CA, USA}, 2015.

\bibitem[Ledig et~al.(2017)Ledig, Theis, Huszar, Caballero, Aitken, Tejani, Totz, Wang, and Shi]{ledig2017photorealistic}
C. Ledig, L. Theis, F. Huszar, J. Caballero, A.~P. Aitken, A. Tejani, J. Totz, Z. Wang, and W. Shi.
\newblock Photo-realistic single image super-resolution using a generative adversarial network.
\newblock \emph{IEEE/CVF Conf. on Comp. Vision and Patt. Recog. (CVPR)}, pages 105--114, 2017.

\bibitem[Liang et~al.(2021)Liang, Cao, Sun, Zhang, Gool, and Timofte]{Liang2021SwinIRIR}
Jingyun Liang, Jie Cao, Guolei Sun, K. Zhang, Luc~Van Gool, and Radu Timofte.
\newblock Swinir: Image restoration using swin transformer.
\newblock \emph{IEEE/CVF Int. Conf. on Comp. Vision Workshops (ICCVW)}, pages 1833--1844, 2021.

\bibitem[Liang et~al.(2022)Liang, Zeng, and Zhang]{details_or_artifacts}
Jie Liang, Hui Zeng, and Lei Zhang.
\newblock Details or artifacts: A locally discriminative learning approach to realistic image super-resolution.
\newblock In \emph{IEEE/CVF Conf. on Comp. Vision and Patt. Recog. (CVPR)}, pages 5657--5666, 2022.

\bibitem[Lim et~al.(2017)Lim, Son, Kim, Nah, and Lee]{EDSR2017}
B. Lim, S. Son, H. Kim, S. Nah, and K.~M. Lee.
\newblock Enhanced deep residual networks for single image super-resolution.
\newblock In \emph{IEEE/CVF CVPR Workshops}, 2017.

\bibitem[Liu et~al.(2023)Liu, Shao, Wang, and Zhang]{ganinv_LIU2023286}
Fukang Liu, Mingwen Shao, Fan Wang, and Lixu Zhang.
\newblock High-fidelity gan inversion by frequency domain guidance.
\newblock \emph{Computers and Graphics}, 114:\penalty0 286--295, 2023.

\bibitem[Liu et~al.(2018)Liu, Li, and Sun]{Liu2018AttributeAwareFA}
Yunfan Liu, Qi Li, and Zhenan Sun.
\newblock Attribute-aware face aging with wavelet-based generative adversarial networks.
\newblock \emph{2019 IEEE/CVF Conference on Computer Vision and Pattern Recognition (CVPR)}, pages 11869--11878, 2018.

\bibitem[Lugmayr et~al.(2020)Lugmayr, Danelljan, van Gool, and Timofte]{srflow}
A. Lugmayr, M. Danelljan, L. van Gool, and R. Timofte.
\newblock {SRFlow:} learning the super-resolution space with normalizing flow.
\newblock In \emph{European Conf. Comp. Vision (ECCV)}, pages 715--732, 2020.

\bibitem[Lugmayr et~al.(2021)Lugmayr, Danelljan, Timofte, et~al.]{2021_ntire}
Andreas Lugmayr, Martin Danelljan, Radu Timofte, et~al.
\newblock Ntire 2021 {L}earning the super-resolution space challenge.
\newblock In \emph{IEEE/CVF Conf. on Computer Vision and Pattern Recognition Workshops (CVPRW)}, pages 596--612, 2021.

\bibitem[Lugmayr et~al.(2022)Lugmayr, Danelljan, Timofte, et~al.]{2022_ntire}
Andreas Lugmayr, Martin Danelljan, Radu Timofte, et~al.
\newblock Ntire 2022 challenge on learning the super-resolution space.
\newblock In \emph{IEEE/CVF Conf. on Computer Vision and Pattern Recognition Workshops (CVPRW)}, pages 785--796, 2022.

\bibitem[Luo et~al.(2023)Luo, Zhu, Xu, and Liu]{dualformer_luo2023effectiveness}
Xin Luo, Yunan Zhu, Shunxin Xu, and Dong Liu.
\newblock On the effectiveness of spectral discriminators for perceptual quality improvement.
\newblock In \emph{ICCV}, 2023.

\bibitem[Ma et~al.(2017)Ma, Yang, Yang, and Yang]{Ma_NRQM}
Chao Ma, Chih-Yuan Yang, Xiaokang Yang, and Ming-Hsuan Yang.
\newblock Learning a no-reference quality metric for single-image super-resolution.
\newblock \emph{Comput. Vis. Image Underst.}, 158:\penalty0 1--16, 2017.

\bibitem[Ma et~al.(2020)Ma, Rao, Cheng, Chen, Lu, and Zhou]{ma_SPSR}
Cheng Ma, Yongming Rao, Yean Cheng, Ce Chen, Jiwen Lu, and Jie Zhou.
\newblock Structure-preserving super resolution with gradient guidance.
\newblock In \emph{IEEE Conf. on Computer Vision and Pattern Recognition (CVPR)}, 2020.

\bibitem[Martin et~al.(2001)Martin, Fowlkes, Tal, and Malik]{bsd100_cite}
D. Martin, C. Fowlkes, D. Tal, and J. Malik.
\newblock A database of human segmented natural images and its application to evaluating segmentation algorithms and measuring ecological statistics.
\newblock In \emph{IEEE Int. Conf. on Computer Vision. (ICCV)}, pages 416--423 vol.2, 2001.

\bibitem[Mechrez et~al.(2018)Mechrez, Talmi, Shama, and Zelnik-Manor]{Mechrez_contentloss}
Roey Mechrez, Itamar Talmi, Firas Shama, and Lihi Zelnik-Manor.
\newblock Maintaining natural image statistics with the contextual loss.
\newblock In \emph{Asian Conf. Comp. Vision (ACCV)}, page 427–443, 2018.

\bibitem[Mei et~al.(2021)Mei, Fan, and Zhou]{mei_nlsa}
Yiqun Mei, Yuchen Fan, and Yuqian Zhou.
\newblock Image super-resolution with non-local sparse attention.
\newblock In \emph{IEEE/CVF Conf. on Computer Vision and Pattern Recognition (CVPR)}, pages 3516--3525, 2021.

\bibitem[Mittal et~al.(2013)Mittal, Soundararajan, and Bovik]{niqe}
Anish Mittal, Rajiv Soundararajan, and Alan~C. Bovik.
\newblock Making a “completely blind” image quality analyzer.
\newblock \emph{IEEE Signal Processing Letters}, 20\penalty0 (3):\penalty0 209--212, 2013.

\bibitem[Moon et~al.(2023)Moon, Kim, and Park]{wagi_moon2023}
SeungJun Moon, Chaewon Kim, and Gyeong-Moon Park.
\newblock Wa{GI}: Wavelet-based {GAN} inversion for preserving high-frequency image details, 2023.

\bibitem[Niu et~al.(2020)Niu, Wen, Ren, Zhang, Yang, Wang, Zhang, Cao, and Shen]{niu_han}
Ben Niu, Weilei Wen, Wenqi Ren, Xiangde Zhang, Lianping Yang, Shuzhen Wang, Kaihao Zhang, Xiaochun Cao, and Haifeng Shen.
\newblock Single image super-resolution via a holistic attention network.
\newblock In \emph{European Conf. Comp Vis.ion (ECCV)}, page 191–207, 2020.

\bibitem[Park et~al.(2022)Park, Moon, and Cho]{fxsr}
Seung~Ho Park, Young~Su Moon, and Nam~Ik Cho.
\newblock Flexible style image super-resolution using conditional objective.
\newblock \emph{IEEE Access}, 10:\penalty0 9774--9792, 2022.

\bibitem[Park et~al.(2023)Park, Moon, and Cho]{srooe_Park_2023_CVPR}
Seung~Ho Park, Young~Su Moon, and Nam~Ik Cho.
\newblock Perception-oriented single image super-resolution using optimal objective estimation.
\newblock In \emph{Proceedings of the IEEE/CVF Conference on Computer Vision and Pattern Recognition (CVPR)}, pages 1725--1735, 2023.

\bibitem[Phung et~al.(2023)Phung, Dao, and Tran]{wav_diff_Phung_2023_CVPR}
Hao Phung, Quan Dao, and Anh Tran.
\newblock Wavelet diffusion models are fast and scalable image generators.
\newblock In \emph{IEEE/CVF Conf. on Computer Vision and Pattern Recognition (CVPR)}, pages 10199--10208, 2023.

\bibitem[{Rakotonirina} and {Rasoanaivo}(2020)]{esrganplus}
N.~C. {Rakotonirina} and A. {Rasoanaivo}.
\newblock Esrgan+: Further improving enhanced super-resolution generative adversarial network.
\newblock In \emph{IEEE Int. Conf. on Acoust., Speech and Signal Processing (ICASSP)}, pages 3637--3641, 2020.

\bibitem[Rombach et~al.(2022)Rombach, Blattmann, Lorenz, Esser, and Ommer]{compvis_rombach2022high}
Robin Rombach, Andreas Blattmann, Dominik Lorenz, Patrick Esser, and Bjorn Ommer.
\newblock High-resolution image synthesis with latent diffusion models.
\newblock In \emph{IEEE/CVF Conf. on Comp. Vision and Patt. Recog.}, pages 10684--10695, 2022.

\bibitem[Rott~Shaham et~al.(2019)Rott~Shaham, Dekel, and Michaeli]{rottshaham2019singan}
T. Rott~Shaham, Tali Dekel, and Tomer Michaeli.
\newblock Singan: Learning a generative model from a single natural image.
\newblock In \emph{IEEE Int. Conf. on Computer Vision (ICCV)}, 2019.

\bibitem[Saharia et~al.(2022)Saharia, Ho, Chan, Salimans, Fleet, and Norouzi]{sr3_saharia2022image}
Chitwan Saharia, Jonathan Ho, William Chan, Tim Salimans, David~J Fleet, and M. Norouzi.
\newblock Image super-resolution via iterative refinement.
\newblock \emph{IEEE Trans. on Pattern Analysis and Machine Intelligence}, 45\penalty0 (4):\penalty0 4713--4726, 2022.

\bibitem[Sahito et~al.(2019)Sahito, Zhiwen, Ahmed, and Memon]{WIDN_sahito2019wavelet}
F. Sahito, P. Zhiwen, J. Ahmed, and R.~A. Memon.
\newblock Wavelet-integrated deep networks for single image super-resolution.
\newblock \emph{Electronics}, 8:\penalty0 553, 2019.

\bibitem[Schwarz et~al.(2021)Schwarz, Liao, and Geiger]{schwarz2021frequency}
Katja Schwarz, Yiyi Liao, and Andreas Geiger.
\newblock On the frequency bias of generative models.
\newblock \emph{Advances in Neural Information Processing Systems}, 34:\penalty0 18126--18136, 2021.

\bibitem[Shen et~al.(2023)Shen, Yan, Sun, Li, and Pan]{wav_inpainting}
Lili Shen, Jie Yan, Xichun Sun, Beichen Li, and Zhaoqing Pan.
\newblock Wavelet-based self-attention gan with collaborative feature fusion for image inpainting.
\newblock \emph{IEEE Trans. on Emerging Topics in Computational Intelligence}, pages 1--14, 2023.

\bibitem[Shi et~al.(2016)Shi, Caballero, Huszár, Totz, Aitken, Bishop, Rueckert, and Wang]{texture_mathing}
Wenzhe Shi, Jose Caballero, Ferenc Huszár, Johannes Totz, Andrew~P. Aitken, Rob Bishop, Daniel Rueckert, and Zehan Wang.
\newblock Real-time single image and video super-resolution using an efficient sub-pixel convolutional neural network.
\newblock In \emph{IEEE Conf. on Comp. Vision and Pattern Recog. (CVPR)}, pages 1874--1883, 2016.

\bibitem[Tong et~al.(2017)Tong, Li, Liu, and Gao]{tong_densenet}
Tong Tong, Gen Li, Xiejie Liu, and Qinquan Gao.
\newblock Image super-resolution using dense skip connections.
\newblock In \emph{IEEE Int. Conf. on Comp. Vision (ICCV)}, pages 4809--4817, 2017.

\bibitem[Wang et~al.(2020)Wang, Deng, Xu, Chen, and Song]{wang2020multi}
Jianyi Wang, Xin Deng, Mai Xu, Congyong Chen, and Yuhang Song.
\newblock Multi-level wavelet-based generative adversarial network for perceptual quality enhancement of compressed video.
\newblock In \emph{European Conference on Computer Vision}, pages 405--421. Springer, 2020.

\bibitem[Wang et~al.(2018)Wang, Yu, Wu, Gu, Liu, Dong, Qiao, and C.~Loy]{wang2018esrgan}
X. Wang, Ke Yu, S. Wu, J. Gu, Y. Liu, C. Dong, Y. Qiao, and C. C.~Loy.
\newblock {ESRGAN:} enhanced super-resolution generative adversarial networks.
\newblock In \emph{European Conf. on Comp. Vision (ECCV) Workshops}, 2018.

\bibitem[Wu et~al.(2023)Wu, Wang, Liu, Wang, and Zhang]{wav_medical}
Weiwen Wu, Yanyang Wang, Qiegen Liu, Ge Wang, and Jianjia Zhang.
\newblock Wavelet-improved score-based generative model for medical imaging.
\newblock \emph{IEEE Transactions on Medical Imaging}, pages 1--1, 2023.

\bibitem[Xue et~al.(2020)Xue, Qiu, Liu, and Jin]{WRAN_xue2020wavelet}
S. Xue, W. Qiu, Fan Liu, and X. Jin.
\newblock Wavelet-based residual attention network for image super-resolution.
\newblock \emph{Neurocomputing}, 382:\penalty0 116--126, 2020.

\bibitem[Zeyde et~al.(2012)Zeyde, Elad, and Protter]{set14_cite}
Roman Zeyde, Michael Elad, and Matan Protter.
\newblock On single image scale-up using sparse-representations.
\newblock In \emph{Curves and Surfaces}, pages 711--730, Berlin, Heidelberg, 2012. Springer.

\bibitem[Zhang et~al.(2021{\natexlab{a}})Zhang, Long, Wang, Piao, Mei, Yang, and Yin]{zhang2021tsan}
Jiqing Zhang, Chengjiang Long, Yuxin Wang, Haiyin Piao, Haiyang Mei, Xin Yang, and Baocai Yin.
\newblock A two-stage attentive network for single image super-resolution.
\newblock \emph{IEEE Trans. on Circuits and Systems for Video Tech.}, 2021{\natexlab{a}}.

\bibitem[Zhang et~al.(2018{\natexlab{a}})Zhang, Isola, Efros, Shechtman, and Wang]{lpips}
R. Zhang, P. Isola, A.~A. Efros, E. Shechtman, and O. Wang.
\newblock The unreasonable effectiveness of deep features as a perceptual metric.
\newblock In \emph{IEEE/CVF Conf. on Comp. Vision and Patt. Recog. (CVPR)}, pages 586--595, 2018{\natexlab{a}}.

\bibitem[Zhang et~al.(2021{\natexlab{b}})Zhang, Bin, Liu, and Blasch]{wggan_ZHANG2021313}
Ran Zhang, Junchi Bin, Zheng Liu, and Erik Blasch.
\newblock Chapter 13 - wggan: A wavelet-guided generative adversarial network for thermal image translation.
\newblock In \emph{Generative Adversarial Networks for Image-to-Image Translation}, pages 313--327. Academic Press, 2021{\natexlab{b}}.

\bibitem[Zhang et~al.(2021{\natexlab{c}})Zhang, Liu, Dong, and Qiao]{zhang2021ranksrgan}
Wenlong Zhang, Yihao Liu, Chao Dong, and Yu Qiao.
\newblock Ranksrgan: Super resolution generative adversarial networks with learning to rank.
\newblock \emph{IEEE Trans. on Pattern Analysis and Machine Intelligence}, 44\penalty0 (10):\penalty0 7149--7166, 2021{\natexlab{c}}.

\bibitem[Zhang et~al.(2022{\natexlab{a}})Zhang, Zeng, Guo, and Zhang]{zhang_elan}
Xindong Zhang, Hui Zeng, Shi Guo, and Lei Zhang.
\newblock Efficient long-range attention network for image super-resolution.
\newblock In \emph{Euro. Conf. Comp. Vision (ECCV)}, 2022{\natexlab{a}}.

\bibitem[Zhang et~al.(2018{\natexlab{b}})Zhang, Li, Li, Wang, Zhong, and Fu]{RCAN2018}
Y. Zhang, K. Li, Kai Li, L. Wang, B. Zhong, and Y. Fu.
\newblock Image super-resolution using very deep residual channel attention networks.
\newblock In \emph{Euro. Conf. Comp. Vis. (ECCV)}, 2018{\natexlab{b}}.

\bibitem[Zhang et~al.(2018{\natexlab{c}})Zhang, Tian, Kong, Zhong, and Fu]{zhang_res_dense}
Yulun Zhang, Yapeng Tian, Yu Kong, Bineng Zhong, and Yun Fu.
\newblock Residual dense network for image super-resolution.
\newblock In \emph{Proceedings of the IEEE conference on computer vision and pattern recognition}, pages 2472--2481, 2018{\natexlab{c}}.

\bibitem[Zhang et~al.(2022{\natexlab{b}})Zhang, Ji, Hao, and Yao]{PDASR}
Yuehan Zhang, Bo Ji, Jia Hao, and Angela Yao.
\newblock Perception-distortion balanced admm optimization for single-image super-resolution.
\newblock In \emph{European Conf. on Comp. Vision (ECCV)}, 2022{\natexlab{b}}.

\bibitem[Zhou et~al.(2020)Zhou, Deng, Tong, and Gao]{GuidedFreqSep}
Yuanbo Zhou, Wei Deng, Tong Tong, and Qinquan Gao.
\newblock Guided frequency separation network for real-world super-resolution.
\newblock In \emph{IEEE/CVF Conf. on Comp. Vision and Patt. Recog. Workshops (CVPRW)}, pages 1722--1731, 2020.

\end{thebibliography}
}


\end{document}